\pdfoutput=1
\documentclass{article}

\usepackage{arxiv}

\usepackage[utf8]{inputenc} % allow utf-8 input
\usepackage[T1]{fontenc}    % use 8-bit T1 fonts
\usepackage{hyperref}       % hyperlinks
\usepackage{url}            % simple URL typesetting
\usepackage{booktabs}       % professional-quality tables
\usepackage{amsfonts}       % blackboard math symbols
\usepackage{nicefrac}       % compact symbols for 1/2, etc.
\usepackage{microtype}      % microtypography
\usepackage{lipsum}

\usepackage{microtype}%if unwanted, comment out or use option "draft"
\usepackage{graphicx}
\usepackage{caption}
\usepackage{subcaption}
\usepackage{mathpartir}
\usepackage{mathtools}
\usepackage{wrapfig}
\usepackage[dvipsnames]{xcolor}
\usepackage{soul}
\usepackage{xspace}
\hypersetup{
    colorlinks = true,
    citecolor  = blue,
    urlcolor   = blue
}
\usepackage{booktabs}
\usepackage{multirow}
\usepackage{listings}
\usepackage{amsmath}
\usepackage{amsthm}

\title{Reliable State Machines:\\ A Framework for Programming Reliable Cloud Services}

\author{
  Suvam Mukherjee \\ 
  Microsoft Research, India \\
  \texttt{t-sumukh@microsoft.com} \\
   \And
 Nitin John Raj \\
  IIIT-Hyderabad, India\\
  \texttt{nitinjohnraj@gmail.com} \\
   \And
 Krishnan Govindraj \\
  Microsoft Research, India\\
  \texttt{t-krgov@microsoft.com} \\
   \And
 Pantazis Deligiannis \\
  Microsoft Research, India\\
  \texttt{pdeligia@microsoft.com} \\
     \And
 Chandramouleswaran Ravichandran \\
 Microsoft Azure, USA\\
  \texttt{chanravi@microsoft.com} \\
  \And
 Akash Lal \\
 Microsoft Research, India\\
  \texttt{akashl@microsoft.com} \\
   \And
 Aseem Rastogi \\
 Microsoft Research, India\\
  \texttt{aseemr@microsoft.com} \\
  \And
  Raja Krishnaswamy \\
 Microsoft Azure, USA\\
  \texttt{rajak@microsoft.com} \\
}

\newtheorem{theorem}{Theorem}[section]

\theoremstyle{definition}
\newtheorem{definition}{Definition}[section]
\newtheorem{proposition}{Proposition}[section]

\newcommand{\psharp}{\texttt{P\#}}
\newcommand{\csharp}{\texttt{C\#}}
\newcommand{\psharptester}{\texttt{PSharpTester}}
\newcommand{\brsm}{\texttt{bRSM}}
\newcommand{\rsm}{\texttt{sfRSM}}
\newcommand{\krsm}{\texttt{kRSM}}
\newcommand{\sfactor}{\texttt{sfActor}}
\newcommand{\rps}{\texttt{RsmPs}}
\newcommand{\pps}{\texttt{ProdPs}}
\newcommand{\sfwrite}{\texttt{sfWrite}}
\newcommand{\sfnet}{\texttt{sfNetwork}}

\newcommand{\figref}[1]{Figure~\ref{Fi:#1}}
\newcommand{\lstref}[1]{Listing~\ref{Lst:#1}}
\newcommand{\sectref}[1]{Section~\ref{Se:#1}}
\newcommand{\sectrefs}[2]{Sections~\ref{Se:#1} and~\ref{Se:#2}}
\newcommand{\theoref}[1]{Theorem~\ref{thm:#1}}
\newcommand{\propref}[1]{Proposition~\ref{prop:#1}}

\renewcommand{\paragraph}[1]{\vspace{2ex} \noindent \textbf{#1}}
\newcommand{\Omit}[1]{}

\newtheorem{property}{Property}

\begin{document}
\maketitle

% ABSTRACT =================================
Building reliable applications for the cloud is challenging because of unpredictable failures during a program's execution. This paper presents a programming framework called \emph{Reliable State Machines} (RSMs), that offers fault-tolerance by construction. Using our framework, a programmer can built an application as several (possibly distributed) RSMs that communicate with each other via messages, much in the style of actor-based programming. Each RSM is additionally fault-tolerant by design and offers the illusion of being ``always-alive''. An RSM is guaranteed to process each input request exactly once, as one would expect in a failure-free environment. The RSM runtime automatically takes care of persisting state and rehydrating it on a failover.
We present the core syntax and semantics of RSMs, along with a formal proof of
\emph{failure-transparency}.
We provide an implementation of the RSM framework and runtime on the .NET platform for deploying services to Microsoft Azure. We carried out an extensive performance evaluation on micro-benchmarks to show that one can build high-throughput applications with RSMs. We also present a case study where we rewrote a significant part of a production cloud service using RSMs. The resulting service has simpler code and exhibits production-grade performance.
%===========================================

%INTRO======================================
\section{Introduction}
\label{Se:intro}

The industry trend in Cloud Computing is increasingly moving towards
companies building and renting \emph{cloud services} to provide
software solutions to their customers~\cite{industrycloud}. A cloud service in
this context refers to a software application that runs on multiple
machines in the cloud, making use of the available resources -- both
compute and storage -- to offer a scalable service to its
customers. In this paper, we consider the problem of programming
reliable cloud services.

Cloud services are essentially distributed
systems consisting of concurrently running, communicating
processes or \emph{agents}\footnote{We use the term agents in this paper as a
programming construct to distinguish from systems constructs like processes or
physical/virtual machines.}. Agents typically maintain state, and process user requests as they
arrive, which may cause their state to get updated. Consider a word-counting
application: the application receives a stream of words (or strings) as input and
continuously produces output in the form of the highest frequency word
that it has seen so far. Programming such an application for the
single-machine scenario is easy -- the application
maintains a map from words to their frequencies as seen so far, and
for each new word, it updates the map and outputs the word if it is
the new highest frequency word.
%% \nitin{updat\st{e}ing}
%% the map for each new word, and \nitin{outputting?} output the highest frequency word
%% accordingly. \nitin{(Using 'updating' and 'outputting' since the three steps mentioned are 
%%   not in sequence; the first is in parallel to the other two)}
However, to design a more scalable 
application, this map
can be split across multiple distributed agents. More specifically, the
distributed word count application can be designed as follows. A
\emph{main} agent receives input words from clients, and sends
each word
to one of several \emph{counting} agents (based on some
criteria, such as the hash of a word) for processing. Every counting agent 
maintains its own word-frequency map and the local maximum; whenever
the local maximum changes, it sends a message to the \emph{max}
agent. The max agent collates the 
local maxima from all the counting agents and outputs the global
maximum.

%% each machine maintains a
%% subset of the map and the results (highest frequency words on each
%% machine) are later aggregated to produce the global maximum.

%% This naturally requires the application to
%% maintain a map from words to their frequencies seen so far. If the
%% universe of possible words is large, then this map has to potentially
%% be distributed 

A reliable cloud service must be resilient to hardware and software
failures that can cause agents to crash, and to
network failures that can cause message duplications,
reorderings, and drops. To handle crashes in the word-counting
service, the programmer needs to use some form of persistent storage for 
the input stream and the word-frequency maps, and write boilerplate code to
read and write this state, while carefully orchestrating it with the
rest of the computation. The programmer must also handle network
message drops (to avoid missing a word) and duplications (to avoid
counting the same occurrence of a word twice). While some existing
programming frameworks and languages for distributed systems, such as
Orleans~\cite{bernstein2014orleans}, Kafka~\cite{kafkapaper},
Akka~\cite{akka}, Azure Service Fabric~\cite{sf-actors}, among others,
provide the
necessary building blocks of persistent storage, transactions, etc.,
the programmer still has to carefully put
them all together. Thus, an application that is quite simple to
program in the single-machine scenario, quickly becomes a non-trivial
task in the distributed setting.

In this paper, we present a novel programming framework, called
\emph{Reliable State Machines} (RSMs), to program reliable, fault-tolerant 
cloud services. The RSM framework enables a programmer to focus only
on the application-specific logic, while providing resilience against
failures -- both machines and network -- through language design and
runtime. 

At a high-level, RSMs are based on the actor style of programming where an RSM
is the unit of concurrency. An RSM is programmed like a communicating state machine -- the programmer
defines the types of events that the RSM can receive, and
handlers for each event type. Optionally, the programmer
can declare some RSM-local state to be persistent. The event handlers
can manipulate state, send messages to
other RSMs, and create new RSMs. Issues of
orchestrating reads and writes of the persistent state with event
handlers, handling network failures, etc., are left to the RSM runtime. 
The runtime ensures that the effects of an event handler
are committed atomically in an all-or-nothing fashion. This ensures that an RSM
appears to process an input message \textit{exactly once}.
In addition, the runtime
provides a networking module for \emph{exact once} delivery
of messages. 
RSMs are build on top of the $\psharp$ framework \cite{DBLP:conf/pldi/DeligiannisDKLT15}, which 
provides convenient .NET syntax for programming state machines \cite{psharpgithub} and enables
programmers to systematically test their applications against functional
specifications \cite{DBLP:conf/fast/DeligiannisMTCD16}.
We provide an overview of the RSM framework, as well as the programming of
the word-counting application using RSMs in Section~\ref{Se:Overview}.

We formalize the syntax and semantics of RSMs and prove a
\emph{failure transparency} theorem. The theorem states that the
semantics of RSMs that includes runtime failures is a refinement of
the failure-free semantics in terms of the observable behavior of an
RSM. As a result, programmers can program and test
their applications assuming failure-free
semantics, while the failure transparency theorem guarantees the
same behavior even in the presence of runtime
failures. Section~\ref{sec:formal} contains details of our
formalization.

We have developed two different implementations of our framework -- one
on top of the Azure Service Fabric platform \cite{servicefabric} and the other
using Apache Kafka \cite{kafka,kafkapaper} -- demonstrating that the basic
concepts behind RSMs are general and can be implemented on different
platforms (\sectref{Implementation}). Our evaluation (\sectref{evalmicro}) 
shows that performance-wise RSMs are competitive with other production
cloud programming frameworks, even with the additional guarantees of failure
transparency, and it is possible to
build high-throughput applications using RSMs.
To evaluate the
programming and testing experience, we present a case study where we
re-implement an existing production-scale backend service of Microsoft Azure. We show that
the RSM implementation of the service is simple,
easier to reason about, amenable to systematic testing via the
$\psharp$ framework, and meets its scalability requirements (\sectref{CaseStudy}).

\section{Overview}
\label{Se:Overview}

This section presents an overview of the RSM framework. We 
show how to program the word-count example with RSMs,
followed by details of the RSM runtime and failure transparency.
In the rest of the paper, we use \textit{events} and \textit{messages}
interchangeably.

\subsection{Programming and testing the word count example}

As mentioned in Section~\ref{Se:intro}, we design the distributed word-count 
application using three types of RSMs: (a) a main RSM that sets
up other RSMs, receives words from the client, and forwards
them to the word-count RSMs, (b) word-count RSMs that 
maintain the highest frequency word they have \emph{individually} seen so far,
and (c) a max-RSM that aggregates local
maxima from the word-count RSMs, and outputs the global maximum.

%\vspace{0.2cm}
\noindent\begin{minipage}{\textwidth}
\begin{lstlisting}[caption={Main-RSM for the word count example.},label={Lst:WordCountCodeMain},captionpos=b,basicstyle=\scriptsize,escapechar=\@,commentstyle=\color{PineGreen}]
event WordEvent: (word: string);  // Event types with their payloads
event WordFreqEvent: (word: string, freq: int);
event InitEvent: (target: rsmId);

machine MainMachine {
  PersistentDictionary<rsmId, int> WordCountMachines;  // Set of word count machines
  PersistentRegister<rsmId> MaxMachineId;  // The rsmId of the aggregator machine

  start state Init { do Initialize }
  state Receive { on WordEvent do ForwardWord }

  void Initialize () {
    var max_id = create (MaxMachine);  // First create the max machine
    store (MaxMachineId, max_id);  // Store it
    for(var i = 0; i < N; ++i) {  // Create the word count machines
      var id = create(WordCountMachine); store(WordCountMachine[id], 1);
      send (id, new InitEvent (max_id));  // Send the max machine id to each word count machine
    }
    jump (Receive);  // Begin receiving events
  }

  rsmId GetTargetMachine (string s) { return load(WordCountMachines[hash(s) mod N]); }
  
  void ForwardWord (WordEvent e) { send (GetTargetMachine (e.word), e); }  // Forward the event
}
\end{lstlisting}
\end{minipage}

\newcommand\ls\lstinline

Listing~\ref{Lst:WordCountCodeMain} shows the source code for the main-RSM using
an abbreviated $\csharp$-like syntax. RSMs are
programmed as state machines. The programmer
first declares the three event types to use in the program:
\ls{WordEvent}, \ls{WordFreqEvent}, and \ls{InitEvent}, each carrying
the mentioned payloads. Values of type \ls{rsmId} (e.g., used in 
the payload of
\ls{InitEvent}) are RSM instance ids. We will explain the
use of these events as we go along.

The main RSM has two states: \ls{Init} is the start state, and
\ls{Receive} is the state in which it receives the input words. The
machine declares two persistent fields: a
\ls{WordCountMachines} dictionary to maintain the \ls{rsmId} of
each word count RSM, and a \ls{MaxMachineId} for the \ls{rsmId} of the
max machine. Fields declared with the ``\ls{Persistent}'' types denote persistent
local state of the RSM. 
In the \ls{Init} state, the main-RSM creates an instance of max-RSM and
\ls{N} instances of word-count RSMs (using the \ls{create} API), and sends
the \ls{rsmId} of the max-RSM instance to every word count RSM as
payload in the \ls{InitEvent} event (using the \ls{send} API). The persistent fields
are also updated (using \ls{store}). The machine then transitions to
the \ls{Receive} state (using the \ls{jump} API). In the \ls{Receive}
state, when the
machine receives a \ls{WordEvent} from the environment, which
contains the next word, it forwards the word to an appropriate max count
machine. Since the \ls{Receive} state specifies no transitions, the
RSM remains in the \ls{Receive} state, ready to receive
the next word.

Listing~\ref{Lst:WordCountCodeWC} shows the code for a word-count RSM.
It maintains, in its persistent state, a running map of word frequencies (\ls{WordFreq}) 
and the highest frequency (\ls{HighFreq})
that it has seen so far. Whenever the highest frequency changes, it
forwards the corresponding word to the max machine, using the
\ls{rsmId} stored in the \ls{TargetMachine} field. This RSM also shows
the use of volatile state in the form of the field
\ls{WordsSeenSinceLastCrash}; this field is reset every time the
RSM fails. Such variables can be used for gathering information
such as program statistics that are not required to survive
failures. Note that the execution of each
handler (its call stack, all local variables, etc.) is also carried
out on volatile memory.

A word count machine has two states: \ls{Init} and
\ls{DoCount}. In the \ls{Init} state, it waits for the \ls{InitEvent}
(from the main machine). The rest of the code is straightforward.

%% When the \texttt{WordCountMachine} RSM is instantiated, it starts in its
%% \textit{start} state \texttt{Init}. It waits for an \texttt{InitEvent},
%% initializes its variables, and transitions to the MachineState \texttt{DoCount}. 
%% Now the RSM waits for messages of type \texttt{WordEvent}. On receiving them, it will execute the method
%% \texttt{Count} that does as expected: it loads the
%% frequency of the received word and increments it. We use \texttt{load} and
%% \texttt{store} to mark accesses to persistent variables. When the highest frequency is
%% exceeded, it sends a message to the \texttt{TargetMachine}.

%% The variable \texttt{WordsSeenSinceLastCrash} is held in volatile memory. 

%\vspace{0.2cm}
\noindent\begin{minipage}{\textwidth}
\begin{lstlisting}[caption={Word count RSM for the word count example.},label={Lst:WordCountCodeWC},captionpos=b,basicstyle=\scriptsize,escapechar=\@,commentstyle=\color{PineGreen}]
machine WordCountMachine {
  PersistentDictionary<string, int> WordFreq;  // Local map for words to their frequencies
  PersistentRegister<int> HighFreq;  // The highest frequency seen so far
  PersistentRegister<rsmId> TargetMachine;  // The max machine rsmId, forwarded by the main machine
  int WordsSeenSinceLastCrash;  // A volatile variable to count words seen since last crash

  start state Init { on InitEvent do Initialize }
  state DoCount { on WordEvent do Count }

  void Initialize (InitEvent e) {  // Wait for the init event from the main machine
    store(TargetMachine, e.target);
    jump (DoCount);
  }

  void Count(WordEvent e) {  // Receive the word from the main machine
    WordsSeenSinceLastCrash++;
    var f = load(WordFreq[e.word]) + 1;  // Increment the frequency of the word by 1 ...
    store(WordFreq[e.word], f);  // And store it back
    if (f > load (HighFreq)) {  // Update the highest frequency, if required
      store (HighFreq, f);
      send(load (TargetMachine), new WordFreqEvent (e.word, f));  // And send it to the max machine
    }
  }
}
\end{lstlisting}
\end{minipage}

The max-RSM, shown below, simply takes a maximum over
the frequencies that it receives, and forwards the maximum one to an
external service (which may print to console or write to an output
file).

%\vspace{0.2cm}
\noindent\begin{minipage}{\textwidth}
\begin{lstlisting}[caption={Max-RSM for the word count example.},label={Lst:WordCountCodeMax},captionpos=b,basicstyle=\scriptsize,escapechar=\@,commentstyle=\color{PineGreen}]
machine MaxMachine {
  PersistentRegister<int> HighFreq;  // Highest frequency seen so far

  state DoCount { on WordFreqEvent do CheckMax }

  void CheckMax(WordFreqEvent e) {  // Update the current highest frequency if needed
    if (e.freq > load (HighFreq)) {
      store (HighFreq, e.freq);
      send (env, e);
    }
  }
}
\end{lstlisting}
\end{minipage}

\paragraph{Implementation.} The RSM programming framework is embedded
in $\csharp$, and uses the $\psharp$ state machine programming
model. Each RSM is defined as a $\csharp$ class, with local-state as
class fields, and event handlers as class methods. Using RSMs does 
not require the user to learn a new programming language. We 
provide more implementation details in Section~\ref{Se:Implementation}.

%% An RSM program begins execution by instantiating the
%% \textit{main} RSM, which can then create additional RSMs. The word counting
%% application can be set up as follows. Suppose a client has multiple
%% words and it wants to compute the highest frequency word among them. If the set
%% of words is large, then storing the word-frequency map in a single RSM may not
%% be efficient.  The main RSM (not shown) can create one instance of \texttt{MaxMachine} 
%% and $N$ instances of the \texttt{WordCountMachine} (for a sufficiently large
%% $N$), each initialized with the \textit{rsmId} of the \texttt{MaxMachine} instance.
%% \textbf{Todo: show a figure}.
 
%% The client will then, for each word, hash it modulo $N$ and send to the
%% corresponding \texttt{WordCountMachine} RSM.

\paragraph{Testing the application.} Having written the application in
our framework, the programmer can also test it by supplying a
specification and ask the $\psharp$ tester to validate the
specification. In the word count application, for example, a
functional correctness specification -- eventually the word with the
highest frequency is output by the \ls{MaxMachine} RSM -- can be
tested. The $\psharp$ tester uses state-of-the-art algorithms to search over the
space of possible executions of an RSM program for bugs \cite{DBLP:conf/sigsoft/DesaiQS15,DBLP:conf/fast/DeligiannisMTCD16}. 
$\psharp$ testing can help catch many bugs. For instance, changing any of 
the persistent variables of the RSMs to volatile will render the
program incorrect; indeed if the \ls{MaxMachine}s don't persist their
word frequency maps, upon restart, their output may not be correct.
If the \ls{MainMachine} does not use the same hash
function inside \ls{GetTargetMachine} for all input words, then also
the specification will fail (because it may forward two
different occurrences of the same word to two different RSMs), etc. We confirmed 
that the $\psharp$ tester is indeed able to  find all these errors very
quickly.

\paragraph{Summary.} Our framework frees the
programmer from the burden of designing and programming for failures. In
the word count application, as we can see above, the
source code \emph{only} contains the application-specific logic,
and no boilerplate code for handling failures, restarts, etc.
There is still concurrency in the program that may be hard to reason about,
which is why we provide $\psharp$ testing. We describe the RSM runtime that provides resilience from
machine and network failures next.

%\footnote{An interesting exercise is to design
%a word-counting application where each of the words output by program are 
%the maximum frequency word in some \textit{prefix} of the word sequence sent out by the
%client. This specification does not hold for the RSM program
%described in this section. We supply a solution in the supplemental
%material.}. 

\subsection{RSM runtime}
%% Aseem: Totally messed up where this was supposed to come ...
%% We demonstrate (\sectref{Implementation}) that the atomic-commit functionality
%% needed for RSMs as well as for exact-once message delivery, 
%% are easily built over existing mature commercial infrastructures, e.g., Azure
%% Service Fabric or Apache Kafka, thus, allowing us to leverage systems-level
%% innovations in this space to design an efficient runtime. The
%% additional mechanism of detecting and resurrecting failed RSMs is also provided by
%% this infrastructure. Our runtime makes it easy to deploy RSM programs to Azure
%% for execution across multiple machines.

\begin{figure}
\centering
\includegraphics[width=0.5\textwidth]{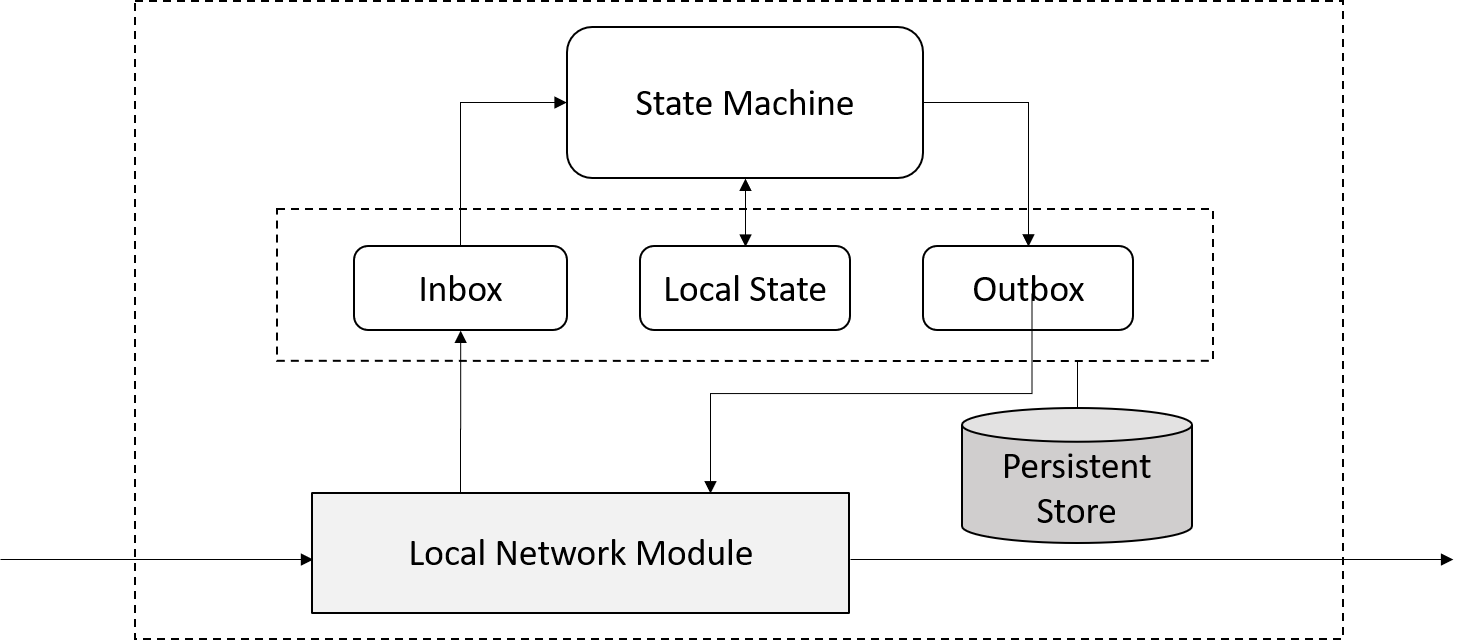}
\caption{Internals of an RSM.}
\label{Fi:RsmInternals}
\end{figure}

%\setlength\intextsep{-0.5pt}
%\begin{wrapfigure}{r}{0\textwidth}
%  %\vspace{-20pt}
%  %\begin{center}
%  \includegraphics[width=0.5\textwidth]{RsmStructure.png}
%\caption{Internals of an RSM.}
%\label{Fi:RsmInternals}
%  %\end{center}
%  %\vspace{-20pt}
%\end{wrapfigure}

\figref{RsmInternals} shows the runtime architecture of a single RSM.
The runtime ensures that each RSM has a unique \ls{rsmId}. An RSM is
associated with its own \textit{inbox} of input events,
an \textit{outbox} of output events (the events that it sends out),
and local state that consists of both persistent and volatile
(in-memory) components. The inbox, outbox, persistent fields, and
the current state of the RSM state machine 
(e.g. \ls{Init} or \ls{Receive} in \lstref{WordCountCodeMain}) 
are backed by a persistent store (e.g. a replicated storage
system). Each RSM also has a local networking module that is
responsible for communicating with other RSMs or to clients or
external services. The inbox and outbox are queues, following the
standard FIFO enqueue and dequeue semantics.

%%
%% Each RSM instance is associated with a globally unique identifier called
%% \textit{rsmId} that other RSMs can use to communicate with it. RSMs expose two
%% key APIs: \textit{CreateRsm}, which instantiates a new RSM of the given type
%% and returns its \textit{rsmId}, and \textit{Send} that sends a message to the 
%% provided destination.
%% We use the term \textit{message} and \textit{event} interchangeably in this
%% paper.

The execution of an RSM consists of three operations. 
\begin{itemize}
\item \textbf{Input}. The networking module receives messages over the
network and enqueues them to the inbox.
\item \textbf{Processing}. The processing inside an RSM is single-threaded.
It iteratively dequeues an event from the inbox 
and processes it by executing its
corresponding event handler. The handler can create other RSMs or send
events to existing ones. Each of these requests are enqueued to the outbox.
The handler can also mutate the persistent and volatile local state of the RSM.
\item \textbf{Output}. The networking module dequeues messages from the outbox
and sends them over the network to their destination.
\end{itemize}
These operations can execute in any order. In our implementation of RSMs
(\sectref{Implementation}), we run
them in parallel using background tasks; we ensure that the  enqueue and 
dequeue operations on the queues (inbox and outbox) are linearizable
\cite{DBLP:journals/toplas/HerlihyW90}, and thus, safe to execute concurrently. 

%% Doing processing entirely on persistent state can be expensive, as well as
%% cumbersome to program. For this purpose, RSMs also have 
%% access to local volatile memory. They can load in persistent state to RAM,
%% perform arbitrary in-memory computation (offering a programming environment that
%% is most familiar to programmers), and then write out the results. 

\paragraph{Exact-once processing.}
The RSM runtime ensures that the effects of an 
event handler are committed to
the persistent storage atomically. In particular,
the dequeue of an event $e$ from the
inbox, and the result of processing $e$ (including all updates made to
persistent fields as well as all enqueues to the outbox)
are committed to the persistent storage in an all-or-nothing
fashion. Thus, if the RSM fails before committing, then on restarting
the RSM, $e$ would still be at the head of the inbox, and none of its effects
would have been propagated to the rest of the system. If 
the RSM fails after committing, then the event $e$ has been
processed and will not appear in the inbox on restart. The RSM only sends out
those events that have been committed successfully to the outbox.

\paragraph{Networking module.} The networking modules work
with each other to ensure exact-once delivery of events between
RSMs, i.e., an event is
dequeued from the outbox of an RSM and enqueued to the inbox of the target RSM
atomically. While exact-once delivery is default, the programmer can
choose more relaxed delivery semantics. All our examples (\sectref{Implementation})
use the stricter exact-once implementation.

To communicate with external non-RSM services, the RSM framework has the 
notion of an \textit{environment} that acts as an interface to the outside world.
The environment can supply input by enqueueing to the inbox of an
RSM. The RSMs can in turn send events to a special \ls{rsmId} called
\ls{env}, which references the environment. Such events 
still get enqueued  to the outbox of the RSM. When committed, they are 
forwarded to their intended  destination through plug-ins to the networking module 
supplied by the user. 

\paragraph{Non-determinism.}
We allow RSM handlers to be non-deterministic, i.e., two executions of an
event handler on the same event and starting from the same local state
may produce different output. For instance, consider an extension to  
the word-count example where each input word is associated with a timestamp. The
main-RSM forwards only those words with a timestamp not older than $24$hours.
Main-RSM can simply look up the current time of day and make the decision of
forwarding the word on not. This action is non-deterministic. 

Non-determinism does not change the RSM guarantees in any way: all
state changes made by an event handler are first committed locally. This ensures
that all non-deterministic choices are resolved and recorded
before they are propagated outside the RSM.

\paragraph{Using $\psharp$ for testing.}
We chose $\psharp$ for two reasons. First, it provides various programming conveniences for writing
state machines and is already in use for writing production 
code \cite{psharpgithub,DBLP:conf/pldi/DesaiGJQRZ13}. Second, $\psharp$ offers means of
writing end-to-end specifications of a
collection of communicating state machines. The specifications (both safety and
liveness) can then be validated using powerful systematic search over the space
of all interleavings of the program. This method has been shown to be
very effective at finding
concurrency bugs
\cite{DBLP:conf/sigsoft/DesaiQS15,DBLP:conf/pldi/DeligiannisDKLT15,DBLP:conf/fast/DeligiannisMTCD16,DBLP:conf/fmcad/MudduluruDDLQ17}. 
In our work, we provide an automatic way of lowering an RSM
program to a $\psharp$ program. A programmer can write the specification
of an RSM program, then validate the specification using
$\psharp$ systematic testing. We provide more details on testing of
our main case study in Section~\ref{Se:caseStudyEval}.

\paragraph{Failure transparency.} Using the exact-once processing and
exact-once delivery, the RSM framework provides a failure transparency
property. The
property essentially says that the observable behavior of an RSM is
independent of the failures of the machine and the network. This
enables the programmers to
focus only on the application-specific logic when programming RSMs,
and also test only for the failure-free executions. The
property relies on the non-interference of the
persistent storage from the volatile class fields. Intuitively, the
volatile class fields are reset on failures, and so, if they leak into
event payloads for example, the crashes can be observable. On the
other hand, volatile local variables of an event handler are
different, as upon a restart, they are always re-initialized.

\section{Formalization of RSMs}
\label{sec:formal}

\newcommand{\kw}[1]{\mbox{\normalfont\lstinline!#1!}}
\newcommand{\core}{\mbox{$R_{SM}$}\xspace}

In this section, we formalize a core of the RSM programming model,
called small RSM or \core, and
its operational semantics. We state and prove the \emph{failure
transparency} theorem in Section~\ref{sec:theorem}.

\subsection{Syntax and semantics}

\begin{figure}[t]
\[
\begin{array}{@{}@{}rlcl}
\text{Field} & f & & \text{Class name} \quad C \quad\quad \text{Integer} \quad n, r\\
\text{Value} & v & ::= & n \mid x \mid f\\
\text{Expression} & e & ::= & v \mid \kw{load}\;f \mid v_1 \oplus v_2 \mid \star\\
\text{Statement} & s & ::= & x := e \mid f := e \mid \kw{store}\;f\;v \mid \kw{if}\;v\;s_1\;s_2 \mid s_1; s_2 \mid \kw{create}\;x\;C \mid \kw{send}\;v_1\;v_2\;v_3\\
\text{Class defn.} & D & ::= & \kw{class}\;C\;\{\overline{\kw{persistent}\;f := n}; \overline{\kw{volatile}\;f := n};\{\overline{x := n}; s\} \}
\end{array}
\]
\caption{\core syntax.}
\label{fig:syntax}
\end{figure}

Figure~\ref{fig:syntax} shows the \core syntax. For simplicity, we
present the syntax in A-normal form~\cite{CPS-reasoning}, where most
of the sub-expressions are values. An RSM $C$ in \core is declared as a
class definition $D$, consisting of \kw{persistent} and \kw{volatile}
fields, and an event handler statement $\{\overline{x := n}; s\}$,
with local variables $\overline{x}$ and statement $s$ (handlers for
specific event types can be encoded in \core using \kw{if}
statements in $s$). All the variables and fields in \core are
integer-typed.

Statements in the language include local variable assignment ($x :=
e$), assignment to \kw{volatile} fields ($f := e$),
\kw{persistent} field updates ($\kw{store}\;f\;v$), conditional
statements ($\kw{if}\;v\;s_1\;s_2$) and sequencing ($s_1; s_2$).
While the \kw{volatile} fields can be
operated upon directly (e.g. adding two of them), to work on the
\kw{persistent} fields, they first need to be \kw{load}ed into local
variables, and then \kw{store}d back.
The form $\kw{create}\;x\;C$ creates a new RSM
$C$ and binds its RSM id to the variable $x$ (the ids are
also integer-valued). Finally the statement form
$\kw{send}\;v_1\;v_2\;v_3$ is used to send an event of event type
$v_2$ (an integer) with payload $v_3$ to the destination machine with
RSM id $v_1$. Expressions $e$ in the language include values $v$,
reading a \kw{persistent} field ($\kw{load}\;f$), and binary
operations $v_1 \oplus v_2$. We also model non-determinism in the
language -- the expression form $\star$ evaluates to a random integer
at runtime.

\subsubsection{Local evaluation judgment}
Operational semantics of \core consists of two judgments, a local
evaluation judgment for reducing the event handler statement to
process an event, and a global judgment where the configuration
consists of all the RSMs executing concurrently. We first present the
local evaluation judgment.

\begin{figure}[t]
\[
\begin{array}{@{}@{}rlcl}
    \text{Field map} & F & ::= & \cdot \mid f \mapsto n, F\\
    \text{Local environment} & L & ::= & \cdot \mid x \mapsto n, L\\
\text{Event list} & E & ::= & \cdot \mid (n_r, n_e, n_p), E\\
\end{array}
\]
\caption{Runtime configuration syntax for local evaluation}
\label{fig:local-runtime-config}
\end{figure}

\begin{figure*}[t]
%%\begin{small}
\fbox{$F; L \vdash e \Downarrow n$}\quad\quad\quad\quad
\fbox{$E; F; L; s \rightarrow E_1; F_1; L_1; s_1$}
\[
\begin{array}{l}
\inferrule*[lab=E-var]
{
}
{
F; L \vdash x \Downarrow L[x]
}
\hspace{0.1cm}
\inferrule*[lab=E-volatile]
{
}
{
F; L \vdash f \Downarrow F[f]
}
\hspace{0.1cm}
\inferrule*[lab=E-persistent]
{
}
{
F; L \vdash \kw{load}\;f \Downarrow F[f]
}
\hspace{0.1cm}
\inferrule*[lab=E-binop]
{
F; L \vdash v_i \Downarrow n_i\\\\
n = n_1 \oplus n_2
}
{
F; L \vdash v_1 \oplus v_2 \Downarrow n
}
\hspace{0.1cm}
\inferrule*[lab=E-star]
{
}
{
F; L \vdash \star \Downarrow n
}\\\\
\inferrule*[lab=L-store]
{
F; L \vdash e \Downarrow n
}
{
E; F; L; \kw{store}\;f\;e \rightarrow E; F[f \mapsto n]; L; \kw{skip}
}
\hspace{0.1cm}
\inferrule*[lab=L-if]
{
F; L \vdash v \Downarrow n\quad
(n \neq 0 \Rightarrow s = s_1) \wedge (n = 0 \Rightarrow s = s_2)
}
{
E; F; L; \kw{if}\;v\;s_1\;s_2 \rightarrow E; F; L; s
}\\\\
\inferrule*[lab=L-create]
{
\mathsf{fresh}\;n_r \quad\quad E_1 = (n_r, n_{C}, 0), E 
}
{
    E; F; L; \kw{create}\;x\;C \rightarrow E_1; F; L[x \mapsto n_r]; \kw{skip}
}
\hspace{0.7cm}
\inferrule*[lab=L-send]
{
F; L \vdash v_i \Downarrow n_i \quad\quad E_1 = (n_1, n_2, n_3), E
}
{
E; F; L; \kw{send}\;v_1\;v_2\;v_3 \rightarrow E_1; F; L; \kw{skip}
}
\end{array}
\]
%%\end{small}
\caption{\core local semantics.}
\label{fig:local-semantics}
\end{figure*}

%% \nitin{Do we need the meet in the second clause of the IF rule?
%% Might read better just to break it into two rules. It's a matter of
%% preference, I suppose.}

Local evaluation judgments are of the form $E; F; L; s \rightarrow E_1; F_1; L_1; s_1$,
where the syntax for $E$, $F$, and $L$ is shown in
Figure~\ref{fig:local-runtime-config}. $F$ and $L$ are field map and
local environment, mapping fields and local variables to values. $L$
contains three special variables $x_s$, $x_e$, and $x_p$ that map to
the source RSM, event type, and the payload of the \emph{current}
event that is being
processed; these fields are initialized in the global judgment. 
%% [\nitin{Two special variables? We don't store the source RSM id? Perhaps we 
%%     don't need to, but I don't see why not.}]
$E$ is
a list of output events. An event is a triple of the form $(r,
n_e, n_p)$, where $r$ is the destination RSM id, $n_e$ is the event
type, and $n_p$ is the event payload. Notably $F$, $L$, and $E$ are
all non-persistent. Their interaction with the persistent state
happens in the global judgment. Statement reduction uses an auxiliary
expression evaluation judgment of the form $F; L \vdash e
\Downarrow n$. Statements at runtime include an
additional \kw{skip} form to denote the terminal statement.

Figure~\ref{fig:local-semantics} shows the selected rules for
statement reduction and expression evaluation. The expression rules
are all standard, notably rule {\sc{E-star}} non-deterministically
evaluates the $\star$ expression to some integer $n$. Most of the
statement reduction rules are also standard. For example, rule
{\sc{L-store}} uses the expression evaluation form to evaluate $e$,
and stores the result in the field map
$F$. Rule {\sc{L-if}} branches based on the evaluated value of $v$.
Rule {\sc{L-create}} simply records the creation request in the output
events list with a special event type $n_C$. Finally, rule
{\sc{L-send}} evaluates each of the \kw{send} arguments, and updates
the output event list $E$.

\subsubsection{Global evaluation judgment}

Global evaluation judgment has the form $S \vdash M; \Pi
\longrightarrow M_1; \Pi_1$. $M$ and $\Pi$ are maps with RSM
ids as domains. The map $M$ maps the RSM ids to local configurations
$E; F; L; s; b$, where $E$, $F$, $L$, and $s$ come from the local
judgment, and $b$ is a (volatile) bit that is $1$ if the machine is
currently processing an event or $0$ otherwise. We will also write
$F_p$ and $F_v$ to denote the $F$ map components for \kw{persistent}
and \kw{volatile} fields respectively. The map $\Pi$ maps
each RSM id to its class $C$ and
persistent storage, i.e. $C; I; O; P; T$, where $I$ is the inbox
persisting the
incoming events, $O$ is the outbox persisting the outgoing events, $P$
is the persistent fields map, and $T$ is the \emph{trace} of the RSM
that records its observable behavior; the trace $T$ is ghost and is
only used to state and prove the failure transparency theorem. The
grammar for $I$, $O$, and $T$ is same as that of the event list $E$,
while persistent field map $P$ is a field map like $F$. Finally, $S$
is the signature that maps class $C$ to its definition.

%\setlength\intextsep{-0.5pt}
%\begin{wrapfigure}{r}{0\textwidth}
%  %\vspace{-20pt}
%  %\begin{center}
%  \includegraphics[width=0.35\textwidth]{statemachine}
%  %\end{center}
%  %\vspace{-20pt}
%\end{wrapfigure}

\begin{figure}
\centering
\includegraphics[width=0.4\textwidth]{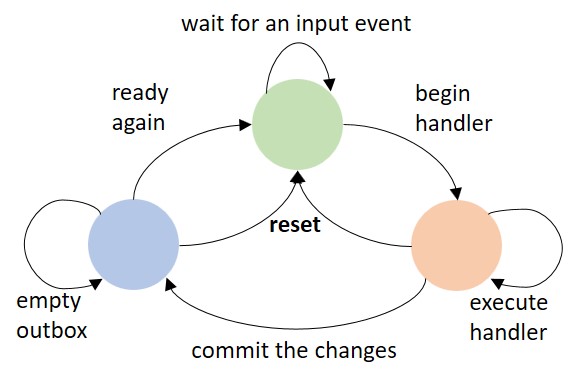}
\caption{Flow of operations for an RSM.}
\end{figure}

At a high-level, each RSM (a) reads an event from its input queue,
(b) processes it using its handler statement, (c) commits the events
generated and the persistent field map in its persistent store, (d)
empties the outbox in the persistent store, and starts from (a)
again. At each of these steps, the machine can crash and recover,
where all of its non-persistent data (including the local state $E$,
$F$, $L$) is lost. The global semantics essentially implements this
state machine for each RSM, while executing the RSMs concurrently with
each other.

%% \aseem{Change the rsm id metavariables to $r$ before this. And is drop
%%   and halt really necessary?}

\begin{figure*}[t]
%%\begin{small}
\[
\begin{array}{l}
\inferrule*[lab=G-start]
{
M(r) = \cdot; F; \_; \kw{skip}; 0\quad
\Pi(r) = C; \_, (n_s, n_e, n_p); \cdot; P; \cdot\\\\
F = F_p \cup F_v \quad F_p = P\\\\
L = \mathsf{initL}(C, n_s, n_e, n_p)\quad
s = \mathsf{handler}(C)
}
{
S \vdash M; \Pi \longrightarrow M[r \mapsto \cdot; F; L; s; 1]; \Pi
}
\hspace{0.3cm}
\inferrule*[lab=G-local]
{
M(r) = E; F; L; s; 1\quad\Pi(r).O = \cdot\\\\
E; F; L; s \rightarrow E_1; F_1; L_1; s_1\\\\
M_1 = M[r \mapsto E_1; F_1; L_1; s_1; 1]
}
{
S \vdash M; \Pi \longrightarrow M_1; \Pi
}
\\\\
\inferrule*[lab=G-commit]
{
M(r) = E; F_p \cup F_v; L; \kw{skip}; 1\quad
\Pi(r) = C; I, \_; \cdot; \_; T\\\\
M_1 = M[r \mapsto \cdot; F_p \cup F_v; L; \kw{skip}; 0]
}
{
S \vdash M; \Pi \longrightarrow M_1; \Pi[r \mapsto C; I; E; F_p; T]
}
\hspace{0.3cm}
\inferrule*[lab=G-create]
{
M(r) = \_; \_; \_; \kw{skip}; 0\\\\
\Pi(r).O = \_, (r_1, n_C, \_)
}
{
S \vdash M; \Pi \longrightarrow M; \mathsf{create}(r, r_1, C, \Pi)
}
\\\\
\inferrule*[lab=G-send]
{
M(r) = \_; \_; \_; \kw{skip}; 0\quad
\Pi(r).O = \_, (r_1, n_e, n_p)
}
{
S \vdash M; \Pi \longrightarrow M; \mathsf{send}(r, r_1, n_e, n_p, \Pi)
}
\hspace{0.3cm}
\inferrule*[lab=G-reset]
{
\Pi(r) = C; \_; \_; P; \_\\\\
M_1 = M[r \mapsto  \cdot; \mathsf{resetF}(C, P); \cdot; \kw{skip}; 0]
}
{
S \vdash M; \Pi \longrightarrow M_1; \Pi
}
%% \inferrule*[lab=L-store]
%% {
%% F; L \vdash e \Downarrow n
%% }
%% {
%% E; F; L; \kw{store}\;f\;e \rightarrow E; F[f \mapsto n]; L; \kw{skip}
%% }
%% \hspace{0.1cm}
%% \inferrule*[lab=L-if]
%% {
%% F; L \vdash v \Downarrow n\\
%% n \neq 0 \Rightarrow s = s_1 \wedge n = 0 \Rightarrow s = s_2
%% }
%% {
%% E; F; L; \kw{if}\;v\;s_1\;s_2 \rightarrow E; F; L; s
%% }\\\\
%% \inferrule*[lab=L-create]
%% {
%% \mathsf{fresh}\;n_r \quad\quad E_1 = (n_r, n_{C}, 0), E
%% }
%% {
%% E; F; L; \kw{create}\;x\;C \rightarrow E_1; F; L; \kw{skip}
%% }
%% \hspace{0.7cm}
%% \inferrule*[lab=L-send]
%% {
%% F; L \vdash v_i \Downarrow n_i \quad\quad O_1 = (n_1, n_2, n_3), O
%% }
%% {
%% E; F; L; \kw{send}\;v_1\;v_2\;v_3 \rightarrow E_1; F; L; \kw{skip}
%% }
\end{array}
\]

\begin{small}
\[
\begin{array}{lcl}
\mathsf{initL}(C, n_s, n_e, n_p) & = & \kw{let}\;S(C) = \kw{class}\;C\;\{\_; \_;\{\overline{x := n}; \_\}\}\;\kw{in}\\
& & (\overline{x \mapsto n}, x_S \mapsto n_s, x_e \mapsto n_e, x_p \mapsto n_p)\\
\mathsf{handler}(C) & = & \kw{let}\;S(C) = \kw{class}\;C\;\{\_; \_;\{\_; s\}\}\;\kw{in}\;s\\
\mathsf{create}(r, r_1, C, \Pi) & = & \kw{let}\;S(C) = \kw{class}\;C\;\{\overline{\kw{persistent}\;f := n}; \_;\_ \}\;\kw{in}\\
& & \kw{let}\;\Pi(r) = C_r; I; O; P; T\;\kw{in}\\
& & \kw{let}\;\Pi_1 = \Pi[r \mapsto C_r; I; \mathsf{tail}\;O; P; (r_1, n_C, 0), T]\;\kw{in}\\
& & \Pi_1[r_1 \mapsto C; \cdot; \cdot; \overline{f \mapsto n}; \cdot]\\
\mathsf{send}(r, r_1, n_e, n_p, \Pi) & = & \kw{let}\;\Pi(r) = C; I; O; P; T\;\kw{in}\;\kw{let}\;\Pi(r_1) = C_1; I_1; O_1; P_1; T_1\;\kw{in}\\
& & \kw{let}\;\Pi_1 = \Pi[r \mapsto C; I; \mathsf{tail}\;O; P; (r_1, n_e, n_p), T]\;\kw{in}\\
& & \Pi_1[r_1 \mapsto C_1; (r, n_e, n_p), I_1; O_1; P_1; T_1]\\
\mathsf{resetF}(C, P) & = & \kw{let}\;S(C) = \kw{class}\;C\;\{\_; \overline{\kw{volatile}\;f := n};\_ \}\;\kw{in}\;P, \overline{f \mapsto n}
\end{array}
\]
\end{small}
%%\end{small}
\caption{\core global semantics.}
\label{fig:global-semantics}
\end{figure*}

%% \nitin{Made small changes to the send macro. ($r, n_e, n_p$) should be added to the back of the queue.}

Figure~\ref{fig:global-semantics} shows the global semantics
judgment. In all the rules, one of the machines $r$ takes the
step. Using the Rule {\sc{G-start}}, a machine $r$ enters the event
handler for processing the head event in the input event queue. The
local state of the machine currently is \emph{at rest}, i.e. $M(r).s =
\kw{skip}$ and $M(r).b = 0$, as well as the outbox $\Pi(r).O$ is
empty. The rule creates the local environment $L$ (using the
$\mathsf{initL}$ auxiliary function, shows in the same figure), by
initializing the local variables as per the RSM definition $S(C)$, and
also adding the mappings for event source, event type and event
payload ($x_s$, $x_e$ and $x_p$). The local state of the machine is
changed to process the
handler statement $s$ and the bit $b$ is set to $1$. The persistent
store $\Pi$ is left unchanged.

Rule {\sc{G-local}} shows the local evaluation rule, where a machine
$r$ takes local step in executing the event handler. The rule uses the
local semantics judgment in the premise, and updates $M(r)$
accordingly.

Once a machine $r$ has finished executing the event handler for the
head input event, it uses the rule {\sc{G-commit}} to commit the
persistent state. In the rule, the local state of the machine
has reached the end of handler execution ($M(r).s = \kw{skip}$ and
$M(r).b = 1$). $M(r)$ is changed by setting the bit $b$ to $0$
and the local event list is reset to empty. The
changes to $\Pi(r)$ are: (a) the head event is removed
from $\Pi(r).I$, (b) the output event list $E$ from the local state is
committed to the outbox $\Pi(r).O$, and (c) the new values of the
\kw{persistent} variables from the local state are committed to
$\Pi(r).P$. The (ghost) trace of the machine $\Pi(r).T$ remains
unchanged; the machine next proceeds to send the events out of the
outbox, and append the trace accordingly.

%% \nitin{Why is the ghost trace set to empty? Doesn't hurt, but it doesn't 
%% seem necessary.}

Rule {\sc{G-create}} handles the create event (rule {\sc{L-create}},
Figure~\ref{fig:local-semantics}). The auxiliary function
$\mathsf{create}$ updates the persistent store $\Pi$. For the creator
machine $r$, it removes the create event from the outbox
$\Pi(r).O$, and adds it to the ghost trace $\Pi(r).T$. For the new
machine $r_1$, it initializes the persistent store by reading off the
initial persistent variables map from the signature $S(C)$.
Rule {\sc{G-send}} sends an event from machine $r$ to $r_1$. The
auxiliary function $\mathsf{send}$ removes the event from the outbox
of $r$, and adds it to the ghost trace, as well as to the inbox of
$r_1$. The rule models the exact-once delivery network module.

Finally, a machine $r$ can fail at any point in the execution. The
rule {\sc{G-reset}} models the machine reset. As expected, upon reset,
the local volatile state, including the event list $E$, \kw{volatile}
variables, environment $L$, are all lost. The fields map in the local
state is re-initialized (using $\mathsf{resetF}$) by reading off the
\kw{persistent} variables from $\Pi(r)$ and \kw{volatile} variables
from the signature $S(C)$. The bit $b$ is also set to $0$. We next
present our main theorem of failure transparency.

\subsection{Failure transparency}
\label{sec:theorem}
To state the theorem, we first define a notion of equivalence for
local states $M(r)$. Below, $r$ is an RSM id.

\begin{definition}[Equivalence of local states]
Two local states, $M_1(r)$ and $M_2(r)$ are equivalent,
written as $M_1(r) \cong M_2(r)$, if they are equal in all components,
except for the volatile class fields in their field maps, i.e.
$M_1(r).E = M_2(r).E$, $M_1(r).F_p = M_2(r).F_p$, $M_1(r).L =
M_2(r).L$, $M_1(r).s = M_2(r).s$, and $M_1(r).b = M_2(r).b$.
\end{definition}

Our failure transparency theorm relies on non-interference of
persistent state from volatile fields. We formally state the property below
(we use $\longrightarrow_r$ to denote the machine $r$ taking a step):

\begin{proposition}[Non-interference]
\label{prop:non-interference}
Let $M; \Pi \longrightarrow_r^* M_1; \Pi$ be a run, s.t. each step in
the run is a {\sc{G-local}} step taken by machine $r$, and $M_1$ is terminal
(i.e. $M_1(r).s = \kw{skip}$). Then, $\forall M'.\;M'(r) \cong M(r)$,
there exists $M_1'$ s.t. $M'; \Pi \longrightarrow_r^* M_1'; \Pi$ where
$M_1'(r) \cong M_1(r)$ and each step is a {\sc{G-local}} step.
\end{proposition}

In a supplementary technical report, we present an information-flow
type system for \core that provides this non-interference property for
well-typed programs. Note that non-determinism in our language does
not raise any complications, since to get this property, we can
essentially \emph{replay} the non-deterministic choices from the run
in the premise to the run in the conclusion.

Given Proposition~\ref{prop:non-interference}, we are now ready to
state the failure transparency theorem. We consider a run of a machine
that processes an event end-to-end. We prove that, given any
such run that includes failures (i.e. the rule {\sc{G-reset}}), we can
construct a run without failures, but with same observable traces
$T$.

\begin{theorem}[Failure transparency]
\label{thm:failure-transparency}
Let $M; \Pi \longrightarrow_r^* M_p; \Pi_p \longrightarrow_r M_c; \Pi_c \longrightarrow_r^* M_1; \Pi_1$, where
$M; \Pi$ is ready for a machine $r$ (i.e. it satisfies the premises of the {\sc{G-start}} rule), and

\begin{enumerate}
\item all steps in $M; \Pi \longrightarrow_r^* M_p; \Pi_p$ are either {\sc{G-start}}, {\sc{G-local}}, or {\sc{G-reset}},
\item $M_p; \Pi_p \longrightarrow_r M_c; \Pi_c$ is a {\sc{G-commit}} step, and
\item all steps in $M_c; \Pi_c \longrightarrow_r^* M_1; \Pi_1$ are either {\sc{G-create}}, {\sc{G-send}}, or {\sc{G-reset}}
\end{enumerate}

Then, $\forall M'. M'(r) \cong M(r)$, there exists $M_1'$ s.t.

\begin{enumerate}
\item $M'; \Pi$ $\longrightarrow_r^* M_1'; \Pi_1$,
\item none of the steps in (a) are {\sc{G-reset}}, and
\item $M_1'(r) \cong M_1(r)$
\end{enumerate}

\end{theorem}

Crucially, $\Pi_1$, and hence the trace of machine $r$ remains same in
the conclusion of the theorem. Thus, we prove that the machine run
with failures is a refinement of the machine run without
failures w.r.t. its observable behavior.

%% \nitin{The section numbering is also off. I think statement reduction should be 3.1.1, global reduction 3.1.2, theorems 3.2. Leaving this as a comment in case that was done intentionally.}

%===========================================

%IMPLEMENTATION=============================
\section{Implementation}
\label{Se:Implementation}

\begin{figure}
\centering
\includegraphics[width=0.75\textwidth]{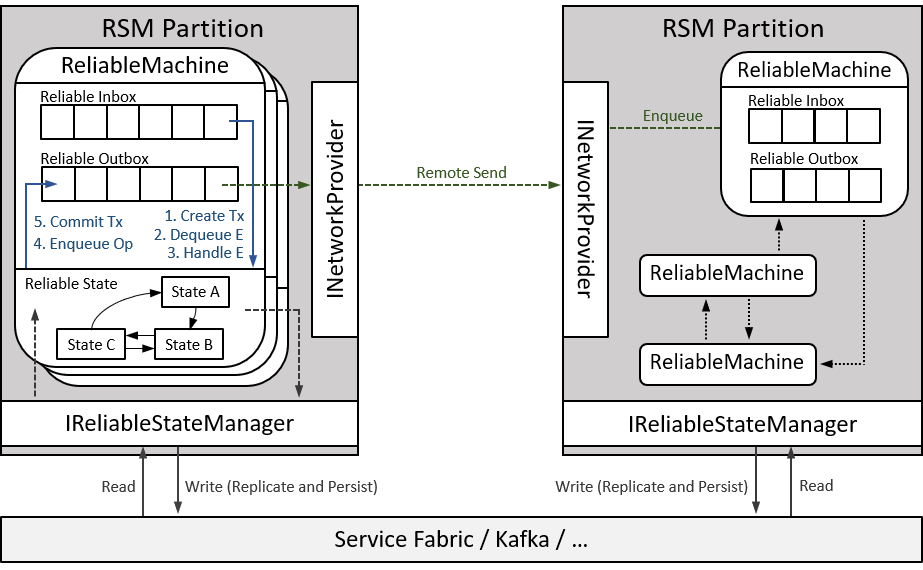}
\caption{The Reliable State Machines implementation.}
\label{Fi:RsmArch}
\end{figure}

This section describes an instantiation of RSMs as a .NET object-oriented
programming framework. The framework is split into two logical parts: 
the \emph{frontend} and the \emph{backend}.
%\setlength\intextsep{-0.5pt}
%\begin{wrapfigure}{r}{0\textwidth}
%  %\vspace{-20pt}
%  %\begin{center}
%  \includegraphics[width=0.65\textwidth]{RSM.png}
%\caption{The Reliable State Machines implementation.}
%\label{Fi:RsmArch}
%  %\end{center}
%  %\vspace{-20pt}
%\end{wrapfigure}
The frontend implements the
programmer-facing APIs while the backend is responsible for the
distributed-system aspects, including state persistence and remote machine communication.
An illustration of the RSM architecture is shown in \figref{RsmArch}.

The frontend exposes an \texttt{RSM.ReliableMachine} base class. An RSM is programmed as
a class that derives from \texttt{ReliableMachine}. An RSM instance is an
object of such a class and event handlers are implemented as methods of the class. 
The base class implements the functionality to drive a 
state machine. The state machine structure is based on $\psharp$, similar to 
the word-count code shown in \sectref{Overview}.  
We focus the discussion here on the reliability aspects of RSMs.
The frontend also provides a runtime, \texttt{RSM.ReliableMachineRuntime},
that implements the APIs for creating RSMs and sending messages
between them. Each RSM
carries a reference to the runtime in order to invoke these APIs. The runtime is
also responsible for \texttt{rsmId} management, ensuring that each RSM is
associated with a unique id throughout its lifetime.

The frontend provides two generic types for declaring local persistent state of 
an RSM: \texttt{RSM.PersistentRegister<T>} and \texttt{RSM.PersistentDictionary<TKey, TValue>}. 
The former implements a \texttt{Get-Put} interface for getting access
to the underlying \texttt{T} object, similar
to the \texttt{load} and \texttt{store} semantics of our formal language. The
object is automatically serialized (on \texttt{Put}) and deserialized (on
\texttt{Get}) in the background.\footnote{We use the \href{https://www.nuget.org/packages/protobuf-net}{\texttt{protobuf-net}} serializer in RSMs, although other 
mechanisms are possible.} The \texttt{PersistentDictionary} type is
similar, although it additionally allows access to individual keys. This has the
advantage that if an RSM handler only accesses a few keys, then only those
keys (and their corresponding values) are serialized and stored, without having 
to serialize the entire dictionary. 
%For this reason, it is more efficient to use
%\texttt{ReliableDictionary<TKey, TValue>} rather than
%\texttt{ReliableRegister< Dictionary<TKey, TValue> >}. (Providing an efficient
%dictionary interface is useful for implementing other kinds of data structures
%and is commonly present in many distributed frameworks such as Orleans,
%ReliableActors, Akka, etc.)

The programmer can declare fields inside an RSM class with these ``\texttt{Persistent}'' 
types to get access to persistent local state. Any other fields in the class 
are treated with volatile semantics. The current state of the state machine
is maintained in a \texttt{PersistentRegister} so that the RSM 
resumes operation from the correct state on failover.

RSMs, once created, stay alive listening to incoming messages, until they are 
explicitly halted. \texttt{ReliableMachine} exposes an option of halting the
RSM. The runtime reclaims any resources of an RSM when it halts.
%Semantically, the halt is equivalent to the RSMs
%becoming zombies (i.e., they drop all incoming messages). The actual runtime
%reclaims the RSM object and its reliable state.

%\textbf{Todo:} A note on other data structures that we provide.

The RSM runtime works against \texttt{RSM.IReliableStateManager} and \texttt{RSM.INetworkProvider} interfaces,
each of which are implemented by the backend. \texttt{IReliableStateManager} is
responsible for creating the inbox and outbox queues, as well as to back the persistent fields of an RSM. 
\texttt{INetworkProvider} allows communication between remote RSMs.
We provide two backend implementations: one using Azure Service Fabric
(\sectrefs{ServiceFabric}{SFOptimizations}) and the other one uses Apache Kafka (\sectref{Kafka}).
We additionally provide a $\psharp$-based backend implementation for the purpose of 
high-coverage systematic testing (\sectref{PSharpBackend}).

\Omit{
\begin{lstlisting}[caption={RSM frontend implementation.},label={Lst:FrontEnd},captionpos=b,basicstyle=\scriptsize,escapechar=\@,commentstyle=\color{PineGreen}]
class ReliableMachine 
{
   // Runtime
   ReliableMachineRuntime runtime;

   // Id
   rsmId Id;

   // Inbox
   IReliableInbox inbox;

   // Outbox
   IReliableOutbox outbox;

   // Event handler loop
   private async Task EventHandlerLoop();
}

class ReliableMachineRuntime
{
   // Set of all hosted RSM Ids
   ReliableDictionary<rsmId, bool> HostedRsmMap;

   // RSM objects
   Dictionary<rsmId, ReliableMachine> IdToRsmMap;

   // CreateRsm
   async Task<rsmId> CreateMachine<T>(Event e);

   // Send
   async Task Send(rsmId r, Event e);
}

\end{lstlisting}
}

%We implemented a fully-featured runtime on top of Azure
%Service Fabric. We also implemented an experimental
%runtime based on Apache Kafka to demonstrate that different backends are
%possible and may offer different performance characteristics. 
%The Kafka backend
%does not provide a full programming experience yet, mostly because the
%.NET bindings of some APIs are not yet available. 

%Each time a new \texttt{ReliableMachine} is created by the
%\texttt{ReliableMachineRuntime}, the \texttt{IReliableStateManager} is
%invoked to provide an instance of a \texttt{IReliableInbox} and a
%\texttt{IReliableOutbox} with the fresh machine. These two queue
%interfaces provide APIs for the reliable inbox and outbox,
%respectively.

%We now provide a brief description of Azure Service Fabric, followed by a
%description of the RSM runtime.

\subsection{Azure Service Fabric backend}
\label{Se:ServiceFabric}

\noindent \textbf{Background.}
Azure Service Fabric (SF)~\cite{servicefabric} provides infrastructure
for designing and deploying
distributed services on Azure. A user begins by setting up an SF cluster
on a required number of Azure VMs. SF sets up a replicated on-disk storage system on the
cluster. An application deployed to an SF cluster benefits from having access to 
co-located storage, instead of having to access a remote storage system.
The store uses primary-secondary-based replication. The user can choose a replication
factor (say, $R$) in which case each update to the store is applied to $R$ replicas, 
with each replica located on a different machine. Updates are only allowed on
primary and then pushed to the secondaries.

SF provides various means of programming a service for deployment to an SF
cluster. The most relevant to our discussion is a stateful application called
\textit{reliable services}~\cite{servicefabricreliable}. 
Such an application consists of multiple 
\textit{partitions}~\cite{servicefabricpartitions}; 
each partition roughly
resembles an individual process constituting the failure domain for the
application. Each partition is associated with its own primary and $R-1$
secondaries. The partition's process is co-located with the primary. 
(Thus, an application with $N$ partitions will have a total of $N$ primaries and
$RN - N$ secondaries, distributed evenly across the SF cluster.)
From the programmer's perspective, each partition gets its own
\texttt{StateManager}~\cite{servicefabricsm} object that provides
access to its store. When a machine carrying a primary fails, one of
its secondaries is promoted to become a primary and the corresponding partition is re-started on
the new primary. A new secondary is elected and brought up to date in the background. Thus, a machine failure
results in restarting of any partition located on it, but all data written to
their \texttt{StateManager} is still available upon restart.

The SF \texttt{StateManager} provides APIs for transacted access
to storage \cite{servicefabrictrans}. A user can 
%% \href{https://docs.microsoft.com/en-us/dotnet/api/microsoft.servicefabric.data.ireliablestatemanager.createtransaction?view=azure-dotnet#Microsoft_ServiceFabric_Data_IReliableStateManager_CreateTransaction}
create
a transaction,
%% (\href{https://docs.microsoft.com/en-us/dotnet/api/microsoft.servicefabric.data.itransaction?view=azure-dotnet}{\texttt{ITransaction}})
use it to 
perform reads and writes to the store, and then commit it. SF
transactions have the database ACID semantics \cite{DBLP:conf/vldb/Gray81}, i.e., they
are atomic, consistent, isolated, and durable with respect to the
other transactions.
As a form of convenience, the user can access the store via a dictionary
interface
%% \href{https://docs.microsoft.com/en-us/dotnet/api/microsoft.servicefabric.data.collections.ireliabledictionary-2?view=azure-dotnet}
(\texttt{IReliableDictionary})
and a queue interface
%% \href{https://docs.microsoft.com/en-us/dotnet/api/microsoft.servicefabric.data.collections.ireliablequeue-1?view=azure-dotnet}
(\texttt{IReliableQueue}). 
These interfaces are shown in \lstref{StateManager}. (We qualify the SF interfaces with \texttt{SF} and
the RSM types with \texttt{RSM} to avoid any confusion.) The
\texttt{SF.IReliableQueue} interface, for example, supports enqueue and dequeue
operations, each of which require the associated transaction. (These are
awaitable $\csharp$ methods \cite{asyncawait}, hence the return type \texttt{Task}.) These operations
appear to take place (with respect to other transactions) only when their associated transaction is committed.
A transaction can span multiple of these reliable collections. The method
\texttt{DictionaryToQueueAtomicTransfer} in \lstref{StateManager} illustrates
an atomic transfer of a value from a dictionary to a queue: it reads from a
dictionary and writes to the queue in the same transaction. 

\noindent\begin{minipage}{\textwidth}
\begin{lstlisting}[caption={Reliable collection interfaces of service fabric
(shown partially) with sample usage.},label={Lst:StateManager},captionpos=b,basicstyle=\scriptsize,escapechar=\@,commentstyle=\color{PineGreen}]
interface SF.IReliableDictionary<TKey, TValue> {
  Task SetAsync(SF.ITransaction, TKey, TValue);
  Task<ConditionalValue<TValue>> TryGetValueAsync(SF.ITransaction, TKey);
}

interface SF.IReliableQueue<T> {
  Task EnqueueAsync(SF.ITransaction, T);
  Task<ConditionalValue<T>> TryDequeueAsync(SF.ITransaction);
}

async void DictionaryToQueueAtomicTransfer(SF.IReliableDictionary<int, int> D,
SF.IReliableQueue<int> Q) 
{
   int key = ...
   using (var tx = StateManager.CreateTransaction()) 
   { 
     var v = await D.TryGetValueAsync(tx, key);
     if(v.HasValue) {
       await Q.EnqueueAsync(tx, v.Value);
     }
     await tx.CommitAsync();
   }
}

\end{lstlisting}
\end{minipage}

\paragraph{RSM backend.}
We can now describe a vanilla implementation of RSMs using SF. Various
optimizations are described in \sectref{SFOptimizations}. An RSM program deploys as a 
stateful service on an SF cluster. A single partition contains exactly one instance of
\texttt{RSM.ReliableMachineRuntime} that may host any number
of RSM instances. 
\texttt{RSM.IReliableStateManager} is implemented as a wrapper on top of the SF
\texttt{StateManager} and the \texttt{RSM.INetworkProvider} on top of 
the SF remoting library for RPC communication \cite{servicefabriccomm}.

The runtime remembers all hosted RSM instances in a persistent dictionary
of the type \texttt{SF.IReliableDictionary<rsmId, bool>}.  
%tracks all hosted Ids and an in-memory map \texttt{IdToRsm} of type 
%\texttt{Dictionary<rsmId, ReliableMachine>} keeps track of the
%RSM objects. 
When a partition comes up (or fails over), it creates a new runtime, which then
immediately reads this dictionary to identify the set of RSMs that it had
hosted before failure (if any). It then re-creates the RSMs with the same ids. All
persistent state associated with an RSM is attached to its id so that an RSM can
rehydrate its state on failover as long as it retains its id.

The types \texttt{RSM.PersistentDictionary} and \texttt{RSM.PersistentRegister} are implemented as
wrappers on top of \texttt{SF.IReliableDictionary}. The RSM types hide SF
transactions from the programmer. The inbox and outbox are just SF reliable queues
(\texttt{SF.IReliableQueue}). An RSM executes as an event-handling loop.
Each iteration of the loop constructs an SF transaction (say,
\texttt{Tx}) and performs a dequeue on the inbox using the transaction. 
If it finds that the queue is empty, the loop terminates and is woken up later
only when a message arrives to the RSM. (This ensures that the RSM takes no
compute resources when it has no work to perform.) If a message is found in the
inbox, then the RSM goes on to execute the corresponding handler. Any access
made by the handler to a persistent field gets attached with the same
transaction \texttt{Tx}. Sending a message $m$ to an RSM $r$ 
is performed as an enqueue of the pair $(m, r)$ to the
outbox queue, also on the same transaction \texttt{Tx}. When the handler
finishes execution, the RSM commits \texttt{Tx} and repeats the 
loop to process other messages in the inbox. Using the same
transaction throughout the lifetime of a handler ensures
that all effects of processing a message happen atomically with the
dequeue of that message.

\paragraph{Networking and exact-once delivery.}
RSMs have two additional background tasks: the first one is responsible for
emptying the outbox, and the other one listens on the network
for incoming messages to add them to the inbox. These tasks are spawned
on-demand as work arrives in order to avoid unnecessary polling. These tasks co-operate to
ensure exact-once delivery between RSMs, even under network failures or delays 
(as long as the connection is eventually established).

The runtime maintains two reliable dictionaries called \texttt{SendCounter} and \texttt{ReceiveCounter}
that map \texttt{rsmId} to \texttt{int}. Pseudo-code for the outbox-draining
task of an RSM with id \texttt{r1} is shown in \lstref{OutboxDraining}. It creates a
transaction \texttt{tx1} and performs a dequeue on the outbox to obtain the 
pair $(m, \texttt{r2})$ of message and destination, respectively. It then sends  
the tuple $(\texttt{r1}, \texttt{SendCounter[r2]}, m)$ over the network to
$\texttt{r2}$ and waits
for an acknowledgement. If it gets the acknowledgement within a certain timeout
period, it increments \texttt{SendCounter[r2]} and commits \texttt{tx1} to complete
the message transfer. If it times-out waiting for an acknowledgement from
$\texttt{r2}$, 
it retries by sending the message again. 

The automatic retry implies that the receiver might get duplicate
messages; however, each such duplicate will be attached with the same counter
value, which the receiver can use for de-duplication. This is achieved in the
input-ingestion procedure shown in \lstref{InputIngestion}. 
The receiver $\texttt{r2}$, when it gets the tuple $(m, c, \texttt{r1})$, first checks if $c$ equals
\texttt{ReceiveCounter[r1]}. If so, it increments \texttt{ReceiveCounter[r1]} and
enqueues $m$ to its inbox. If not, it drops the message because its a
duplicate. Regardless, it always sends an acknowledgement back to $\texttt{r1}$.

\noindent\begin{minipage}{\textwidth}
\begin{lstlisting}[caption={Outbox draining task for RSM $r1$.},label={Lst:OutboxDraining},captionpos=b,basicstyle=\scriptsize,escapechar=\@,commentstyle=\color{PineGreen}]
do:
  create transaction tx1
  (m, r2) = Outbox.Dequeue(tx1);
  c = SendCounter[r2].Get(tx1);
  SendCounter[r2].Put(c + 1, tx1);
  do:
     send (m, c, r1) to r2
  repeat until an ack is received within timeout
  commit tx1
repeat forever
\end{lstlisting}
\end{minipage}

\noindent\begin{minipage}{\textwidth}
\begin{lstlisting}[caption={Input ingestion procedure for RSM
$r2$.},label={Lst:InputIngestion},captionpos=b,basicstyle=\scriptsize,escapechar=\@,commentstyle=\color{PineGreen}]
On receiving (m, c, r1):
  create transaction tx2
  d = ReceiveCounter[r1].Get(tx2);
  if d == c then:
     ReceiveCounter[r1].Put(d+1, tx2);
     inbox.Enqueue(m, tx2);
  send ack back to r1;
  commit tx2
\end{lstlisting}
\end{minipage}

Note that each of the tasks including input-ingestion, outbox-draining, and the
event-handling, use their own transactions that are different from each other. 
This enables the RSM to run these tasks completely independently and
in parallel to each other. SF transactions provide ACID semantics, 
so concurrent enqueue and dequeue operations on queues are safe. 

\paragraph{RSM creation.} When an RSM \texttt{r1} wishes to instantiate a new RSM of
class $C$, it
first creates a globally unique \texttt{rsmId} $r$. This creation can
be done in
several ways. Our implementation uses inter-partition communication to 
first decide the partition that will host the newly created RSM. 
It then grabs a unique counter value from that partition. The pair of partition
name and unique counter value on that partition makes the \texttt{rsmId}
globally unique. Once this value $r$ is obtained, \texttt{r1} enqueues the pair
$(r, C)$  to its outbox. No RSM is actually created until the pair is committed to the
outbox: only its id is constructed eagerly. If
\texttt{r1} fails before committing, then the value $r$ is lost forever. When
\texttt{r1} is restarted, it will construct a new (but still globally unique) id. 

The outbox-draining task of \texttt{r1}, when it picks up a tuple $(r, C)$, will send a message to 
the partition on which $r$ is located. Like before, this message is sent
repeatedly until acknowledged. On the receipt of this message, the RSM runtime 
instantiates a new RSM of type $C$ \textit{only if} it does not already have 
an RSM associated with $r$. If it does have such an RSM, then it drops the
message because it must be a duplicate request, one that it has carried
out already. The recipient sends back an acknowledgement to the sender
regardless.

\subsection{Optimizing the SF backend}
\label{Se:SFOptimizations}

The following lists some of the most important performance optimizations that we
found useful for the SF backend. 

\paragraph{Shared inbox and outbox.}
Creating a separate reliable queue for the inbox and outbox of each RSM does
not scale well unfortunately, especially when the application creates a large number of
RSMs. Each creation incurs an I/O operation. To optimize the RSM creation
time, we instead use a single data structure that is shared across all RSMs in the same partition: 
one for all inboxes and one for all outboxes. 

These shared structures are implemented as an \texttt{SF.IReliableDictionary}
whose key is a tuple of \texttt{rsmId} and an index (\texttt{long}). Each RSM
maintains its own head and tail indices, denoting the contiguous index range that contains
its inbox or outbox contents.  An RSM \texttt{r1}, for instance, can enqueue $m$ to its
outbox by writing it to the key $(\texttt{r1}, \text{tail})$ and incrementing
$\text{tail}$. For efficiency, the head and tail values are only kept in-memory.
On failover, the RSM runtime reads through the shared dictionary to
identify the per-RSM head and tail values, before it instantiates the RSMs with
these values. Additional care is required to ensure proper synchronized access
to head and tail values by the various tasks associated with an RSM. Using these shared structures 
allowed us to significantly scale machine creations (\sectref{micro}).

\paragraph{Batching.} 
We use batching in various forms to optimize overall throughput
(\sectref{micro}). First, the
event-handling loop of an RSM can dequeue multiple messages from its inbox in
the same transaction and process all of them (sequentially, one after the other) before committing all
of their effects together. The commit is a high-latency operation because SF
must replicate all updates to the secondaries and wait for a quorum. This form
of inbox-batching helps hide some of this latency. Second, the outbox-draining
task can dequeue multiple messages from the outbox in the same transaction, and as long as they are
intended for the same destination partition, send them over the network together as a batch.

\paragraph{Non-persistent inbox.} Sending a message $m$ from RSM \texttt{r1} to
\texttt{r2} requires several I/O operations: \texttt{r1} first commits $m$ to its outbox, next it sends $m$
over the network to \texttt{r2}, and finally \texttt{r2} commits $m$ to its inbox.
Interestingly, we can do away with a persistent inbox and only keep it in memory 
without sacrificing any of
the RSM framework guarantees. Our optimization works as follows. The
input-ingestion task of \texttt{r2} simply enqueues $m$ to an in-memory inbox but it
does not immediately send an acknowledgement back to \texttt{r1}. Instead, \texttt{r2} waits
until it is done processing $m$. After \texttt{r2} commits the effects of processing $m$
to its own outbox, it sends the acknowledgement back to \texttt{r1}, after which
\texttt{r1} will remove $m$ from
its outbox. This is safe: the message sits in the (persistent) outbox of \texttt{r1} until 
\texttt{r2} is done processing it. 

%When we do a send, we enqueue to an
%in-memory non-reliable inbox and wait for an acknowledgement from the
%target machine that it has processed this enqueued event. This
%acknowledgement is sent only after the event handler that processed
%this event has committed. If this operation times out, then the sender
%machine retries to send the same event (deduplication logic guarantees
%that the same event will not be processed twice). This process
%effectively guarantees that the event will be eventually processed
%exactly once and will not be lost upon a failure.

%To avoid increasing latency, if this send and wait operation is
%expected to take a long time (e.g. because the target machine has
%multiple events in its inbox that are waiting processing, based on a
%cost model), we enqueue instead to the original reliable
%inbox. Careful synchronization makes sure that the reliable and
%in-memory inbox can co-exist.

\subsection{Kafka backend}
\label{Se:Kafka}

Apache Kafka \cite{kafka,kafkapaper} is a popular distributed messaging platform
that has been used in large production systems by companies such as Netflix, Pinterest and Spotify
\cite{kafka-poweredby}. Kafka supports named sequences of messages 
called \emph{topics}, each being persisted and replicated for fault-tolerance. 
A \emph{producer} appends messages to the tail of a topic. There is no explicit
deletion of messages; the user can configure an expiry time, upon which the
corresponding messages are removed by the system. In order to read a message, a \emph{consumer} subscribes to the topic and
maintains a per-topic index, referred to as the consumer's \emph{offset}. The read cycle 
involves the consumer reading the message at its offset,
incrementing the offset, and then storing the new offset value in a topic of its
own called the \textit{offset-topic}. Kafka supports different consumers to read from
different offsets of a topic concurrently. Starting in version
v0.11.0, Kafka introduced the notion of cross-topic \textit{transactions}. These 
allow a producer to write to multiple topics atomically: either all
the writes succeed or none of them do. Consumers cannot observe
the writes made in a transaction until the transaction commits.  A Kafka
\textit{stream} is a combination of a Kafka producer and consumer: it
consumes messages from an input topic and publishes messages to one or
more output topics. Kafka supports building stateful applications on
top of streams via a key-value state store and convenient Java/Scale
APIs. Exact-once processing of messages can be achieved by transactionally
writing the offset, state and published messages to their respective
topics.

\paragraph{Kafka-based RSMs.} At a high level, the Kafka-based backend for RSMs
comprises of the following: Kafka topics serve the role of persistent queues,
Kafka transactions allow exact-once processing, and Kafka streams provide APIs
for a key-value store to back an RSM's persistent local state. We put these
various features of Kafka together to support the RSM semantics. 

A Kafka RSM (K-RSM) has an associated \emph{inbox topic}, and a \emph{state
topic} for its persistent local state. The RSM also maintains its read offset
into the inbox as part of its persistent local state. An RSM executes as
follows: it reads a message $m$ from the inbox at its read offset and starts a Kafka
transaction \texttt{Tx}. It then runs the handler code for $m$. 
Any changes to the persistent local state are written to the state
topic under \texttt{Tx}. Any message sends are written directly to the inbox topics 
of the receiver K-RSMs, also under \texttt{Tx}. Finally, the incremented offset
is written to the state topic and the transaction \texttt{Tx} is committed. Note
that there was no need to have an \textit{outbox}: Kafka transactions ensure
that the effects of processing a message by one RSM are not observed by other RSMs until
its transaction commits. (SF transactions, on the other hand, cannot span across
reliable collections in different partitions, which is why we needed an outbox
for the SF backend.) 
Restart of an RSM simply involves recovering its state from the state topic that
additionally provides it the read offset of the last un-processed message.

%\setlength\intextsep{-0.5pt}
%\begin{wrapfigure}{r}{0\textwidth}
%  %\vspace{-20pt}
%  %\begin{center}
%  \includegraphics[width=0.5\textwidth]{RSM-Kafka.png}
%  \caption{\label{Fi:krsm}RSM runtime based on Kafka.}
%  %\end{center}
%  %\vspace{-20pt}
%\end{wrapfigure}

A user begins by starting a Kafka cluster, configured to their own
requirements. The K-RSM backend then attaches to the cluster to execute the RSM
program. Unlike SF reliable collections, Kafka topics must be
preallocated to a fixed number, which would typically be much smaller than the
number of RSM instances that a program may create. The K-RSM backend shares a single topic across multiple
RSM instances, which works because each RSM maintains its own offset value. 
The assignment of RSMs to topics is currently 
done in a simple round-robin fashion but more
sophisticated policies are possible as well. Similar to the SF backend, 
messaging in Kafka benefits greatly from batching: both when writing to
a destination topic and when reading from the inbox topic. 

\subsection{$\psharp$ backend}
\label{Se:PSharpBackend}

We additionally designed a backend for the purpose of testing RSM programs. The
backend does not support distribution; it simulates the entire program execution in a
single process. The backend essentially translates an RSM program to a $\psharp$ program for systematic
testing against a specification. We first briefly summarize $\psharp$
capabilities \cite{DBLP:conf/pldi/DeligiannisDKLT15}.

$\psharp$ provides an in-memory framework for implementing concurrent programs; it does not
provide any support for distribution or persistence. A $\psharp$
program consists of multiple state machines that communicate via messages. The
$\psharptester$ tool takes a $\psharp$ program as input and repeatedly executes
it multiple times. It takes over the scheduling of the program so that it can
search over the space of all possible interleavings. $\psharptester$
employs a state-of-the-art portfolio of search strategies that has proven to be
effective in finding bugs quickly
\cite{DBLP:conf/sigsoft/DesaiQS15,DBLP:conf/fast/DeligiannisMTCD16}.
A user can write a specification in the form of a monitor that is checked by the
$\psharptester$ in each execution of the program. Both safety and liveness
specifications \cite{DBLP:conf/fmcad/MudduluruDDLQ17} are supported. 

The $\psharp$ backend for RSMs allows one to write specification monitors in the
same way as $\psharp$ and test their correctness using $\psharptester$. It is
worth noting that the backend is designed with the intention of testing the
user logic as opposed to the RSM runtime itself. For this, the backend
ensures that only the concurrency (and complexity) 
in the user program is exposed to the $\psharptester$; the concurrency 
inside the runtime (which is useful for gaining performance) is disabled.

An RSM translates almost directly to a $\psharp$ machine, with the following
modifications. First, the backend provides mock implementations for all 
persistent types (simulated in-memory for efficiency). Second, 
the three tasks associated with an RSM (i.e., input-ingestion, 
event-handling and outbox-draining) are run sequentially, one after the other.
Third, the exact-once network delivery algorithm is assumed correct, so the
outbox-to-inbox transfer is done atomically (and in-memory). 

An important aspect of the backend is simulating failures in the RSM program. The
failure-transparency property of RSMs crucially helps here: as long as the programmer
makes correct use of volatile memory (\propref{non-interference}), 
failures have no effect at all on the semantics of the program
(\theoref{failure-transparency}). Thus, the backend only needs to check
for \propref{non-interference} on the program. This is done as follows. The backend,
at the time it is about to commit a transaction in the event-handling loop of an
RSM, non-deterministically chooses to carry out the following steps: (1) record the persistent state of the
RSM (both local state and outbox), (2) reset the volatile state of the RSM, 
(3) abort the transaction, thus requiring the RSM to re-process the input
message, and (4) when the RSM reaches the commit point again, assert that
the persistent state equals the recorded state. If a failure of this assertion
is reported by $\psharptester$, the programmer is directly informed of incorrect
usage of volatile state.

%===========================================

%CASESTUDY==================================
\section{Case-Study: PoolServer}
\label{Se:CaseStudy}

We used the RSM framework to redesign the core functionality of an in-production
service, called the \textit{PoolServer}, on Microsoft Azure. This section describes the operations
supported by the service (\sectref{ps-desc}) and its implementation using
RSMs (\sectref{ps-rsm}), highlighting the gains in programmability and testing
of the service. We demonstrate scalability of the RSM code in \sectref{caseStudyEval}.

\begin{figure}[ht!]
        \centering
        \begin{subfigure}[b]{0.35\textwidth}
                \raggedleft
                \includegraphics[width=0.9\textwidth]{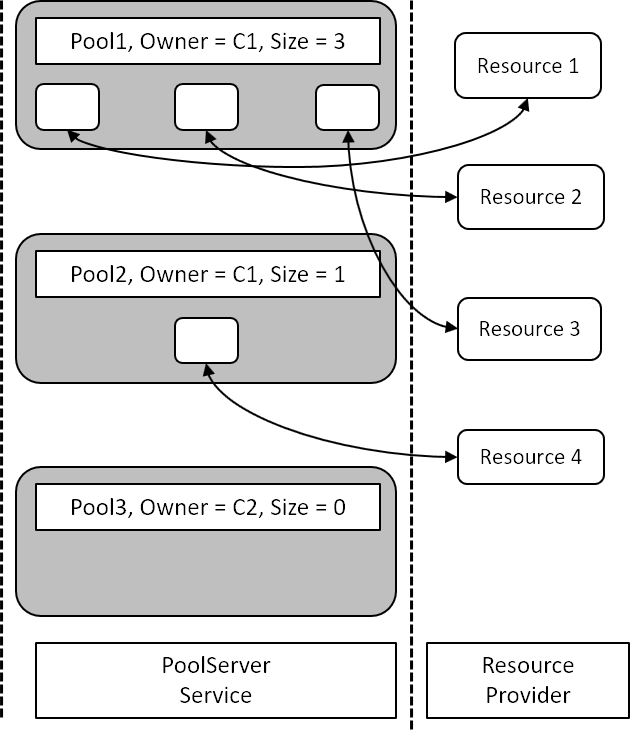}
                \caption{Overview of the PoolServer microservice.}
                \label{fig:ps-overview}
        \end{subfigure} \qquad
        \begin{subfigure}[b]{0.5\textwidth}
                \raggedright
                \includegraphics[width=\textwidth]{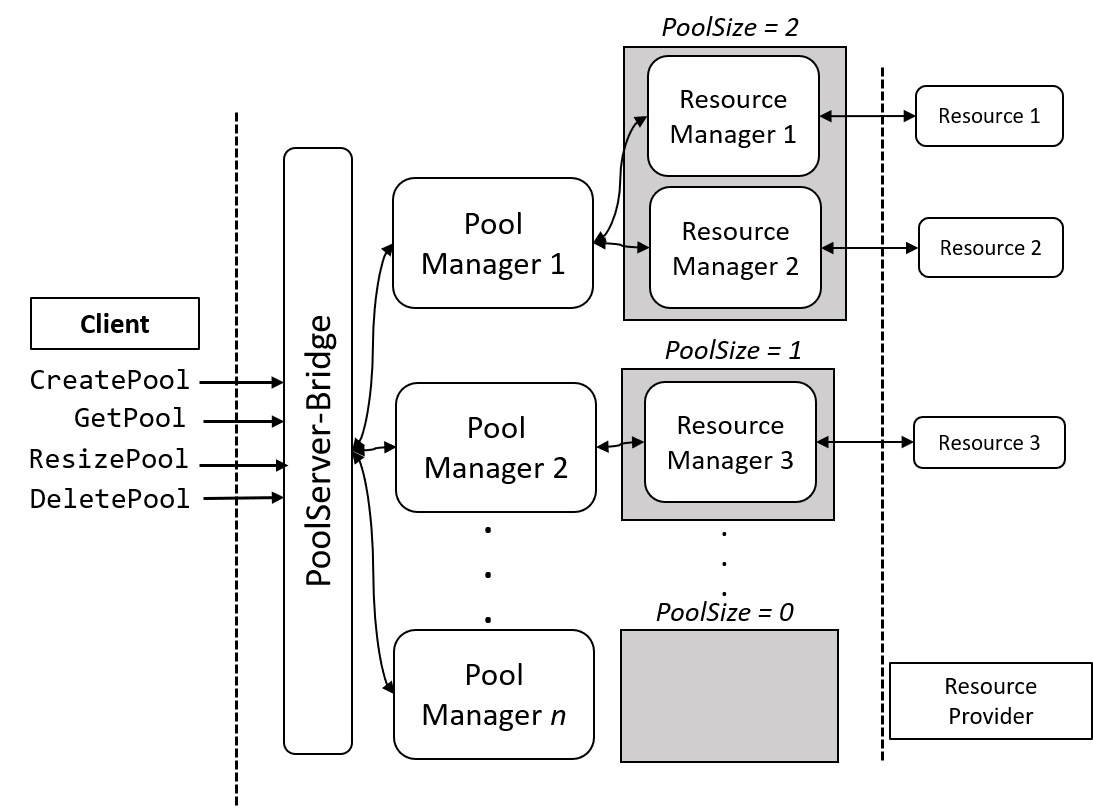}
                \caption{RSM based implementation of PoolServer.}
                \label{fig:ps}
        \end{subfigure}%
        \caption{Achitecture of the PoolServer service.}\label{fig:ps-desc}
\end{figure}

\subsection{Service description}
\label{Se:ps-desc}

The PoolServer (PS) is a generic resource management service. A cloud platform
will typically provide various kinds of compute and storage resources, 
for instance, virtual machines, that can be used in conjunction by a user to
implement certain functionality. The PoolServer is designed to offer 
a convenient abstraction over a low-level resource provider to maintain a
collection of resources. A user can request the PoolServer for a set of $n$
resources (called a \textit{pool}).  The PoolServer calls into the resource
provider to allocate these resources. 

Fig.~\ref{fig:ps-overview} shows a high-level view of the PoolServer
(PS). Each pool has a designated \emph{owner} and
supervises a number of resources. Individual resources can turn
\textit{unhealthy} (e.g., a VM becomes unresponsive), in which case, it is the responsibility of PS to explicitly
delete that resource and allocate a new one to ensure that each pool eventually
reaches its desired size. Also, there should be no \textit{garbage} resources: one that is
allocated by the resource provider but is not associated with any pool. 

A client $C$ can fire a pool creation request to PS, with the desired
number of resources $n$ as a parameter. In response, PS creates a
fresh pool, owned by $C$, with $n$ resources in it. The client can
query the health, resize or delete any existing pool that it owns.

The PoolServer must be responsive and scalable. It must be able to handle pool
creation requests from multiple clients at the same time. Further, the creation
of a pool itself should not add much overhead over the actual allocation of the
resources. PS should also tolerate failures. If the PS crashes, 
it should not lose information about the pools that it had already created, or was in the
middle of creating. For instance, if a requested pool of size $10$ had reached
size $3$ when the PS crashed, it must resume and allocate the remaining $7$. 

\subsection{RSM based PoolServer}
\label{Se:ps-rsm}

We implemented the PoolServer using RSMs. We denote this implementation as
$\rps$. It supports the core functionality that was described in the previous section. In
comparison, the real production service (denoted $\pps$) offers a richer API to
its clients, but the additional features are unrelated to matters of reliability or concurrency. 
Fig.~\ref{fig:ps} shows the high-level
architecture of \rps. There are two RSM types: one called the 
\textit{resource manager} (RM) that is responsible for the lifetime
of a single resource, and another called \textit{pool manager} (PM) that is responsible
for the lifetime of a single pool. This division ensures that the complexities
of dealing with the external resource provider are limited to the RM.  Future changes 
to the resource provider APIs will likely not impact the PM. 

\begin{figure}[ht!]
        \includegraphics[width=\textwidth]{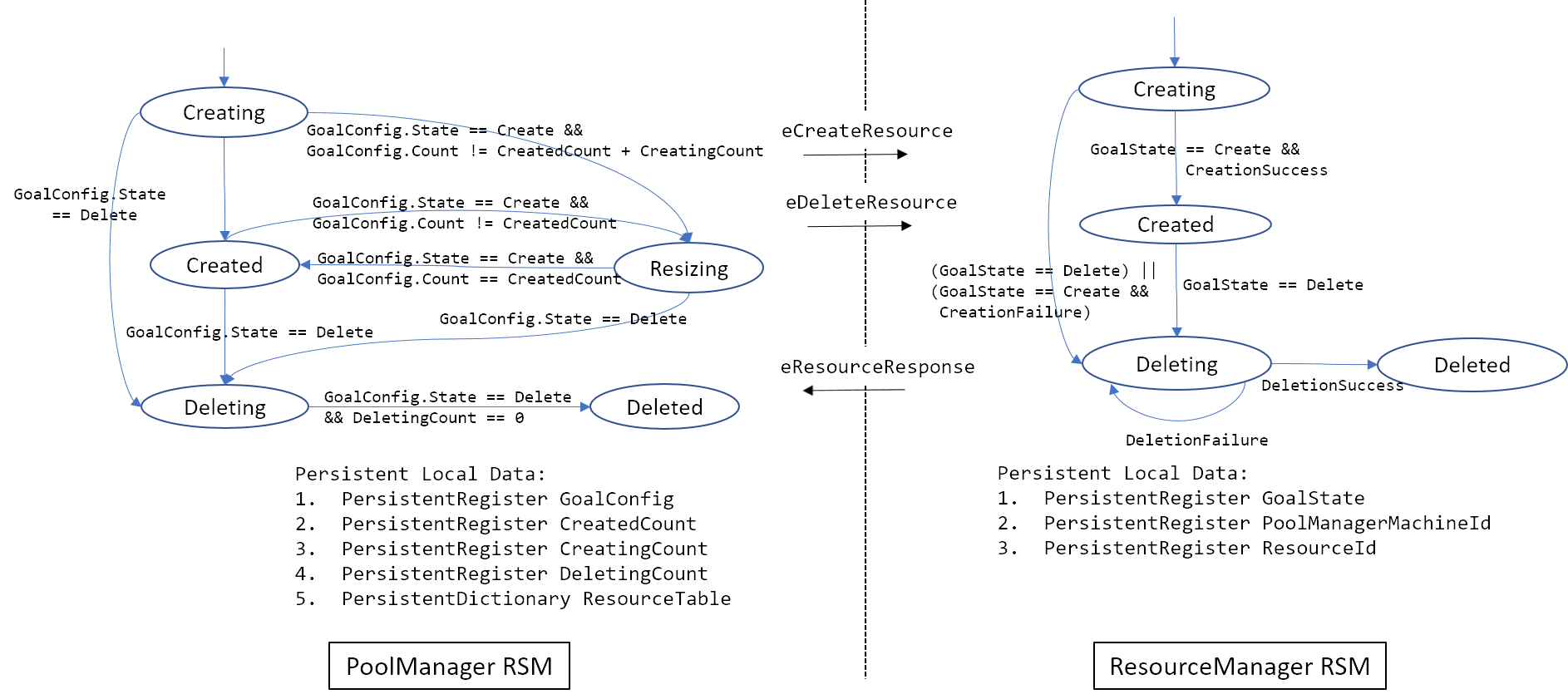} 
        \caption{The pool manager and resource manager RSM state machines.}
        \label{Fi:PS-rsm}
\end{figure}

A client can issue requests such as \texttt{CreatePool}, \texttt{GetPool}, 
\texttt{ResizePool} or \texttt{DeletePool} to $\rps$. These requests are
translated to messages that are directed to the PM that owns the
corresponding pool. The state machine structures of RM and PM are shown
pictorially in \figref{PS-rsm}. We explain the functioning of these RSMs by
tracing through the \texttt{CreatePool} operation.

In response to a client's pool creation request, $\rps$ 
creates a new PM instance. Each such instance maintains three counters: 
\texttt{CreatingCount}, \texttt{CreatedCount} and \texttt{DeletingCount} which are,
respectively, the number of resources that are under creation,
already created, and under deletion. The PM additionally
maintains a \texttt{GoalConfig} that specifies the desired number
of resources in the pool (\texttt{Count}), and the intended
\texttt{State} of the pool (either \texttt{Create} or
\texttt{Delete}). Finally, the RSM maintains a dictionary \texttt{ResourceTable}
containing the \texttt{rsmId}s of all the RM instances that it owns. 

A PM instance starts off in the \texttt{Creating} state with an empty
\texttt{ResourceTable} and each counter set to $0$. Its \texttt{GoalConfig} will get
initialized to the pool size that was requested by the client (on receiving the creation request) 
and the RSM will transition to its \texttt{Resizing} state realizing that it does not have enough resources
created. In the resizing state, the RSM looks at the difference between
$\texttt{GoalState.Count}$ and $\texttt{CreatingCount} + \texttt{CreatedCount}$,
say $m$, and fires off the operation \texttt{ScaleUp(pmId, m)} whose code 
is shown in \lstref{scaleUp}, where $\texttt{pmId}$ is the \texttt{rsmId} of the
current PM instance. We note that this entire operation is devoid of any failover or
retry logic: the PM does not have to worry about failures of the machine hosting
it, or about the failures of the RM instances that it creates. The runtime ensures that the
exact number of instances requested will be eventually created (and no more).

\noindent\begin{minipage}{\textwidth}
\begin{lstlisting}[caption={\texttt{ScaleUp} operation to create resources in a
pool.},label={Lst:scaleUp},captionpos=b,basicstyle=\scriptsize,escapechar=\@,commentstyle=\color{PineGreen}]
void ScaleUp(RsmId pmId, int toCreate)
{
  for (int i = 0; i < toCreate; i++) {
    // Start off an RM to allocate a fresh resource.
    var id = create(ResourceManager);
    send(id, eCreateResource(pmId, ResourceGoalState.Create));
    // Record the creation in the resource table, and we're done.
    store(ResourceTable[id], ResourceState.Creating);
    store(CreatingCount, (load CreatingCount) + 1);
  }
}
\end{lstlisting}
\end{minipage}

\begin{sloppy}
An RM instance reliably persists the handle ($\texttt{PoolManagerMachineId}$) 
to the PM instance that created it, the goal state ($\texttt{GloalState}$) that is
either \texttt{Create} or \texttt{Delete}, and the resource identifier
(\texttt{ResourceId}) returned by the resource provider. 
An RM starts off in the \texttt{Creating} state, fires off a request to the
resource provider, which if successful (\texttt{CreationSuccess}) causes a 
transition to the \texttt{Created} state. It then informs the PM about
successful creation of the resource. The PM waits in its
\texttt{Resizing} state until it gets enough success responses from its RM
instances, i.e., until $\texttt{GoalConfig.Count == CreatedCount}$.
\end{sloppy}

If a resource ever goes unhealthy, the corresponding RM instance transitions 
to the \texttt{Deleting} state and asks the resource provider 
to de-allocate the resource. On successful deallocation, 
the RM transitions to the \texttt{Deleted} state, and informs the PM,
upon which the PM will issue the \texttt{ScaleUp} operation to allocate a
new resource.
Pool deletion is similar and implemented via a corresponding \texttt{ScaleDown}
operation. Both RMs and PMs halt themselves after transitioning to the
\texttt{Deleted} state. 

\paragraph{Correctness.} We use the $\psharp$-testing backend to
check the conformance of $\rps$ to the following specifications. The
testing helped weed out several bugs while implementing the RSM program.
These properties were tested against a model of the resource provider where the
allocation of a resource can non-deterministically fail (but eventually
allocation is successful on repeated attempts) and the resource can go
unhealthy at any time.

\begin{property}
\label{prop:scaling}
Immediately following a \texttt{ScaleUp} or \texttt{ScaleDown}
operation, the number of resources under creation, or
already created, equals the desired number of resources.
\end{property}

\begin{property}
\label{prop:poolSize}
If a client issues the sequence of requests
$\mathtt{CreatePool}(n_1)$, $\mathtt{ResizePool}(n_2)$, $\ldots$,
$\mathtt{ResizePool}(n_k)$, then $\rps$ will \textit{eventually} create a pool
with exactly $n_k$ resources.
\end{property}

\begin{property}
\label{prop:deletePool}
On issuing a \texttt{DeletePool}, eventually all resources of the pool are disposed.
\end{property}

\paragraph{A comparison of $\rps$ with $\pps$.} 
The resource and pool managers lend themselves naturally to a state
machine encoding. The state machines manage the life-cycle of a resource or a
pool, respectively. $\pps$ had a similar design, however, communication was not
through message passing but rather via shared tables, maintained as SF reliable
collections. One agent would update a table and other agents would 
continuously pool these tables to get the updates. Polling
increased CPU utilization: $\rps$ uses roughly $10\times$ less CPU than $\pps$. 
Implicit communication also made the code harder to reason for correctness. 

A direct comparison between the code size of $\pps$ and $\rps$ is not possible
because the former implements more features. However, $\rps$ implements all of
the core functionality in approximately $2000$ lines of code, several times
smaller than the corresponding functionality in $\pps$. The designers of $\pps$
attest to the benefits listed here.

To contain code
complexity, $\pps$ was not designed to be responsive during resize operations:
it would wait to finish one resize operation before looking at subsequent resize
requests. $\rps$, on the other hand, is fully responsive in such scenarios. The PM state machine can
handle new resize requests while it is in the \texttt{Resize} state: it simply
updates its \texttt{GoalConfig} and issues either \texttt{ScaleUp} and \texttt{ScaleDown} until the pool
reaches its goal state. Importantly, the $\psharp$-based testing infrastructure of RSMs
provides strong confidence in exploring a more responsive (and more complex) state-machine design. 
We show in \sectref{caseStudyEval} that $\rps$ is able to comfortably attain production scales. 
%===========================================

%EVALUATION=================================
\section{Evaluation}
\label{Se:evalmicro}

This section reports on a performance evaluation of our RSM implementation. 
\sectref{micro} measures common performance metrics on
micro-benchmarks. \sectref{caseStudyEval} evaluates the performance of our
implementation of the PoolServer case study ($\rps$).
We draw comparisons with the
\href{https://docs.microsoft.com/en-us/azure/service-fabric/service-fabric-reliable-actors-introduction}{Reliable
Actors} 
programming model of Service Fabric \cite{sf-actors} (denoted $\sfactor$). Reliable actors are an implementation of the ``virtual
actors'' paradigm \cite{bernstein2014orleans}. It serves as a useful baseline
for experimentation because it builds on SF much like our SF backend
implementation. Further, reliable actors do not provide failure transparency guarantees, although the
programmer is given access to a persistent key-value store. This allows us to measure the
relative overheads with providing a by-construction fault-tolerant runtime.
In the rest of this section, we use the generic term \emph{agents} to
denote both $\sfactor$s and RSMs.

\subsection{Microbenchmarks}
\label{Se:micro}
Our microbenchmarks evaluate the following three scenarios: $(i)$
\emph{creation}: where we measure the creation time for agents, 
$(ii)$ \emph{messaging latency} between two agents and $(iii)$
\emph{processing throughput}, where we measure the time taken to
process a sequence of messages by an agent. In the subsequent discussion, we use
$\rsm$ and $\brsm$ to denote the SF-based RSM implementation,
with and without optimizations mentioned in \sectref{SFOptimizations},
respectively. We use $\krsm$ to denote the Kafka-based RSM implementation.

\textit{Cluster Setup.} The $\sfactor$, $\brsm$ and $\rsm$ services were
deployed on a $5$-node Service Fabric cluster on Microsoft Azure,
where each node had a $\texttt{D4\_v2}$ configuration ($8$ CPU cores, $28$GB RAM,
and a $400$GB local solid-state drive). The Kafka experiments were run
on an Azure HDInsight cluster with the following configuration: $(i)$ 2
\emph{head} nodes of type $\texttt{D4\_v2}$ executing the RSM runtime and
application $(ii)$ 3 \emph{worker} nodes hosting the Kafka topics, with
a total of $24$ cores and $84$GB RAM, and a total of $6$ premium disks
of size $1$TB each $(iii)$ $3$ nodes for running Apache Zookeeper, with a total of $12$
cores and $21$GB RAM. (Zookeeper serves as a 
coordinator for Kafka nodes and manages cluster metadata.) Because of the
different cluster setup, $\rsm$ and $\krsm$ are not directly comparable.

\paragraph{Creation.} 
It is
important to keep overheads with creation low in order to provide most flexibility in
programming RSM applications. In this experiment, we measure the time taken by a
client to sequentially create $n$ agents. Both the client and the created agents 
reside on the same partition, which allows us to eliminate any networking
overheads from the creation times. 

\begin{figure}
\centering
\includegraphics[width=0.6\textwidth]{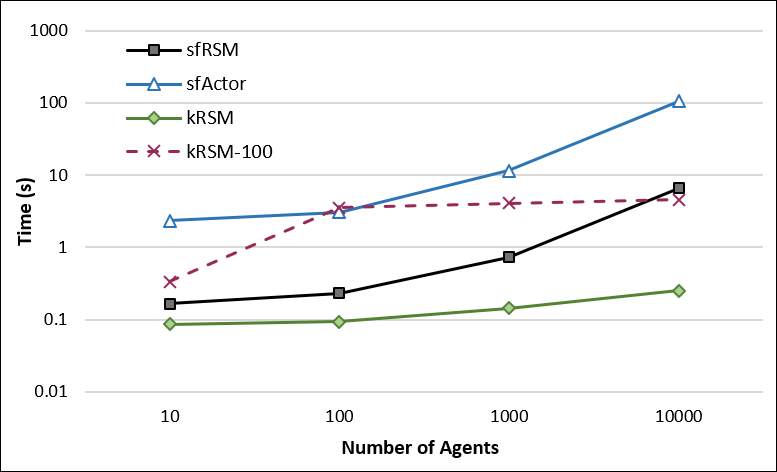}
\caption{$\sfactor$ and $\rsm$ creation times.}
  \label{fig:creation-varyingagents}
\end{figure}

Fig.~\ref{fig:creation-varyingagents} summarizes the
results. The average creation time for $\rsm$ is $5.1$ms, nearly $14X$ faster than $\sfactor$ ($71.4$ms). With all optimizations turned off, the
average creation time of $\brsm$ ($22.2$ms) is $4.4X$ that of $\rsm$. The speedup in creation time for $\rsm$ primarily stems from the shared inbox-outbox optimization. The creation times for both $\sfactor$ and
$\rsm$ scale linearly with the number $n$ of agents created. For $\rsm$, the bulk of the creation time is expended in committing a single SF transaction, which persists the initial local state of the machine and its $\texttt{rsmId}$ to the runtime. Creations in $\krsm$ are measured differently. We create the 
Kafka topics ahead of time because $(i)$ creating topics on-the-fly is much
slower than pre-creating them in bulk, and more importantly $(ii)$ 
there is a limit to the number of topics that can be supported on each
worker node. A $\krsm$ creation now simply involves assigning two existing topics from the pool, along with persisting the id and initial state. We run two experiments, $\krsm$-$1$ and $\krsm$-$100$, where we multiplex the $n$ RSMs onto a single topic and $100$ topics, respectively. Note that all the writes during creation for $\krsm$-$1$ are batched into a single transaction, while the writes for $\krsm$-$100$ involve $100$ transactions. As Fig.~\ref{fig:creation-varyingagents} shows, both $\krsm$-$1$ and $\krsm$-$100$ creations are fairly lightweight, 
with average creation times of $2.4$ms and $2.6$ms respectively (but discounting
the topic pre-creation time).

%\setlength\intextsep{-0.5pt}
%\begin{wrapfigure}{r}{0\textwidth}
%  %\vspace{-20pt}
%  %\begin{center}
%  \includegraphics[width=0.6\textwidth]{Creation.png}
%  \caption{$\sfactor$ and $\rsm$ creation times.}
%  \label{fig:creation-varyingagents}
%  %\end{center}
%  %\vspace{-20pt}
%\end{wrapfigure}

In a separate experiment, we measured the creation
throughput, by firing $1000$ creation requests in parallel. $\sfactor$ and $\brsm$ could achieve
a maximum throughput of $287$ and $67$ creations per second, respectively,
while $\rsm$ could hit a maximum of $1189$ creations per second. The faster
creations for $\rsm$ stems from its optimizations, which result in frugal CPU
and IO requirements.  $\krsm$ creation throughput was $6661$ creations per
second.

%The creation times
%increase at the rate $\mathcal{O}(n)$, where $n$ is the number of $\krsm$s. This
%is due to the increased commits which reliably store metadata like the
%machine-id and topic-id for the new $\krsm$. However, with the optimization of
%sharing topics between $\krsm$, the creation time essentially remains constant.
%The reason for this is that topic sharing allows us share a single transaction
%for all the created $\krsm$s, thereby allowing us to batch all the updates into
%a single commit.

\paragraph{Messaging.}
This experiment measures the cost of exact-once messaging. The experiment
comprises of two agents that repeatedly send a single message ($50$
bytes) back-and-forth and we measure messaging latencies. Messaging in $\sfactor$ is unreliable
(best-effort, and lost on failures). We optionally make the agents in
$\sfactor$ persist their incoming message.

\begin{table}[ht!]
\centering
    \begin{tabular}{@{}|c|c|c|c|c|@{}}
\hline
\textbf{Framework} & \textbf{0.5} & \textbf{0.9} & \textbf{0.99} & \textbf{Mean} \\ \hline \hline
$\sfactor$ & 4.5 & 8 & 9.8 & 4.5 \\ \hline
$\sfactor$\textbf{-Persist} & 12.5 & 23 & 23.5 & 11.9 \\ \hline
$\rsm$ & 23 & 31.5 & 70.6 & 22.8 \\ \hline
$\krsm$ & 8.8 & 10 & 13.5 & 9.1 \\ \hline
\end{tabular}
\caption{\label{tab:mlatency}Messaging latencies.}
\end{table}

Table.~\ref{tab:mlatency} shows the latency measurements at different quantiles.
Unsurprisingly, $\sfactor$ exhibits the lowest latencies. 
When we persist the messages in $\sfactor$, which introduces one write per
message transfer, it increases the latency significantly. $\rsm$ 
requires two write operations per message, making it nearly twice as expensive
as $\sfactor$ with message persistence. Kafka, being a messaging system, is
optimized for low-latency operations, even with exact-once guarantees. $\krsm$
has better latency than $\sfactor$ with message persistence.

\paragraph{Throughput.}
In this experiment, a producer and a consumer agent are located on different
partitions. The producer keeps sending messages (with a varying payload size) to the
consumer. The consumer simply keeps a running count of the number of bytes
received.  We measure the time taken to 
process all the requests, and report the throughput in
MB/s. 

\begin{figure}[ht!]
\centering
 \includegraphics[width=0.6\textwidth]{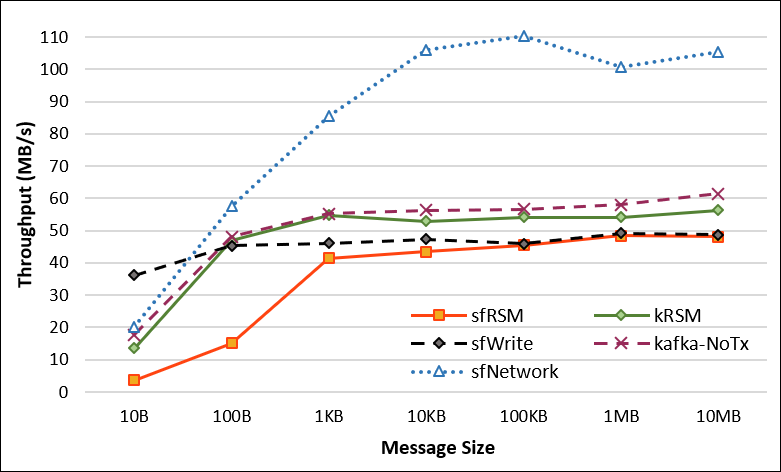}
  \caption{\label{fi:txput}$\rsm$ and theoretical throughputs.}
\end{figure}

Fig.~\ref{fi:txput} summarizes the results. $\rsm$ automatically batches messages
to increase throughput. $\sfactor$ has no default batching mechanism, although increasing
message sizes decreases benefits to be gained from batching. At large message sizes,
$\sfactor$ was able to achieve a maximum throughput of $86.6$MB/s. In comparison, the
maximum throughput for $\rsm$ (across all message sizes) was $48.4$MB/s. 
To account for this difference, we precisely timed all micro-operations involved in the $\rsm$ runtime.

%\setlength\intextsep{-0.5pt}
%\begin{wrapfigure}{r}{0\textwidth}
%  %\vspace{-20pt}
%  %\begin{center}
%  \includegraphics[width=0.6\textwidth]{Txput.png}
%  \caption{\label{fi:txput}$\rsm$ and theoritical throughputs.}
%  %\end{center}
%  %\vspace{-20pt}
%\end{wrapfigure}

Sending a message from the producer to the consumer at least involves writing to the
outbox and then sending the message over the network. We separately measured the
best throughput of writing to a SF reliable collection ($\sfwrite$) and sending
data over the network as fast as possible via (unreliable) RPC ($\sfnet$).
Clearly, the throughput of $\rsm$ will be bounded by the smaller of these two
values. 
As Fig.~\ref{fi:txput} shows, the writes constitute the limiting factor, and 
$\rsm$ incurs very little overhead over the $\sfwrite$ throughput, especially for
large message sizes. Smaller message sizes implies a larger number of messages
per batch, which increases the serialization overhead, and the number of
times the consumer executes its handler. This effect, consequently, widens the gap
between the $\rsm$ and $\sfwrite$ throughputs for smaller message sizes. This
result shows that any improvements in the write throughput of reliable
collections will directly speed up RSMs. The gap between $\rsm$ and $\sfnet$ is
the cost of reliable messaging. Nonetheless, even at the small message size of
$100$bytes, $\rsm$ are able to do roughly $150K$ message transfers per second;
enough for many realistic applications.

$\krsm$ throughput peaks at $56.2$MB/s. With Kafka, the persistence and message
transfer happen together as a topic write. The upper bound for
$\krsm$ is to use non-transactional writes (\texttt{kafka-NoTx}).
Fig.~\ref{fi:txput} shows that $\krsm$ have little overhead compared to the throughput of 
\texttt{kafka-NoTx}.

%perform slightly better (although on a different cluster). Kafka as the backend. The
%theoretical throughput limit is now Kafka non-transactional writes which, as we
%see in Fig.~\ref{fi:txput}, is faster than $\sfwrite$. Thus, the throughput
%exhibited by $\krsm$ ($56.2$MB/s maximum), which employs transactional writes,
%typically exceeds $\sfwrite$ (and therefore $\rsm$).

\subsection{$\rps$ Case Study}
\label{Se:caseStudyEval}

\begin{figure}[ht!]
\centering
 \includegraphics[width=0.6\textwidth]{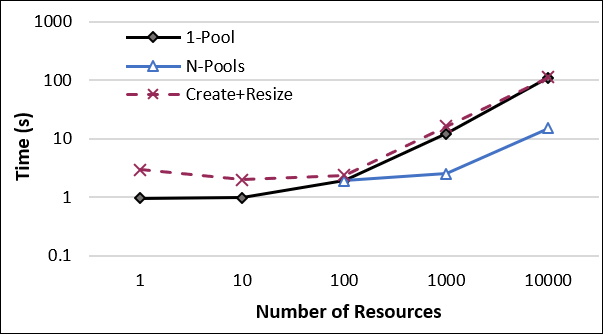}
  \caption{\label{fi:prsm-scale}$\rps$ resource creation.}
\end{figure}

\noindent \textbf{Performance. }
We measure the time taken to create a given number of resources in a
single partition, assuming that the resource-provider calls are instantaneous.
Fig.~\ref{fi:prsm-scale} summarizes the results.

%\setlength\intextsep{-0.5pt}
%\begin{wrapfigure}{r}{0\textwidth}
%  %\vspace{-20pt}
%  %\begin{center}
%  \includegraphics[width=0.6\textwidth]{CreatePool.png}
%  \caption{\label{fi:prsm-scale}$\rps$ resource creation.}
%  %\end{center}
%  %\vspace{-20pt}
%\end{wrapfigure}

In the first experiment, denoted as $1$-Pool, we create a \emph{single} pool
with progressively increasing number of resources. The more realistic scenario,
which arises in production, is to have \emph{multiple} pools of small sizes in a
single partition. In the N-Pools experiment, we create multiple pools (each of
size $100$) in parallel such that the total number of resources matches the
$X$-axis. We make two observations: (i) the creation times for both $1$-Pool and
N-Pools increase linearly with the number of resources (ii) for the same number
of resources, the increased parallelism in N-Pools results in the creation times
being an order of magnitude faster than $1$-Pool. We would like to emphasize
that the workloads here are realistic, and are based on requirements provided by
the developers of the in-production $\pps$ service. The aforementioned results
were reviewed by the developers, who confirmed that $\rps$ comfortably scales to
production workloads. 

To evaluate the responsiveness of $\rps$, we issue $\texttt{CreatePool}(y)$,
followed immediately by $\texttt{ResizePool}(x)$, where $x = y / 100$. The
requirement is to ensure that the total time stays close to
$\texttt{CreatePool}(x)$. The Create+Resize line in Fig.~\ref{fi:prsm-scale}
summarizes the result (with the value $x$ on the $X$-axis). We see that as we increase $x$, the Create+Resize curve
lies very close to $1$-Pool, which is testament to the service's responsiveness.
For small values of $x$, the gap is wider because almost all of the $y$ allocations 
kick-in by the time the resize request is processed.

\paragraph{Testing. }For testing, we create mocks of both the client and the
Resource Provider services, since they are external to $\rps$. Our mocks are
vanilla $\psharp$ machines. The testing exercise was done on a laptop with a
dual-core i$7$ processor, with $8$GB RAM. The tester performed $100$ iterations,
with a scheduling strategy chosen from a pre-defined portfolio, with each
exploration having a depth of $10,000$ steps. Note that the test for
Property~\ref{prop:scaling} is a safety-check, while the tests for
Properties~\ref{prop:poolSize} and \ref{prop:deletePool} are liveness-checks.
The client issued a $\texttt{CreatePool}(100)$ request. We deliberately injected
a bug in the $\texttt{ScaleUp}$ operation by removing the updates to
$\texttt{CreatingCount}$. The resulting violation of
Property~\ref{prop:scaling} was detecting in $0.75$s, generating an error
witness of around $64$ steps. We fixed the error, and issued
$\texttt{CreatePool}(100)$, followed by $\texttt{ResizePool}(5)$, and
Property~\ref{prop:poolSize} was verified in $147.9$s. To verify
Property~\ref{prop:deletePool}, we issue $\texttt{CreatePool}(50)$ followed by a
deletion and the tester verified the property in $119.1$s.

We further injected a bug by converting the $\texttt{CreatedCount}$ to be
volatile. (This means that if the machine was in the middle of a creation operation when it failed, 
it would lose track of all the resources it had created, and therefore the pool would never reach the
$\texttt{Created}$ state.) The tester is able to quickly find 
a violation of property~\ref{prop:poolSize}, in $5.5$s.

%Optional
\paragraph{Other applications. }We have evaluated the applicability of the RSM
language and runtime by encoding several other real-world applications. One
example is a Banking application, where accounts are encoded as RSMs, and there
are \emph{broker} RSMs which are tasked with transferring money from one account
to the other, without incurring any financial losses on failures. The
application specification can be encoded as a liveness property. Another example is a Survey application
\cite{DBLP:conf/popl/RamalingamV13, tailspin}, where \emph{subscribers} can
create surveys, which users can respond to. Each survey
is managed by an RSM, and an overall coordinator RSM creates surveys, reports
survey status, deletes surveys, etc. From a user perspective, responsiveness is
a key metric. The application also needs to ensure specifications like a user
vote is counted exactly once. The RSM framework allowed us to design 
these responsive applications, with all the specifications thoroughly tested.
%===========================================

%RELATED====================================
\section{Related Work}
\label{Se:related}

\noindent \textbf{Actor frameworks.}
Actor-based programming \cite{DBLP:reference/parallel/KarmaniA11} refers to the
general style of programming where the concurrent entities in the program (called
\textit{actors}) each have their own local state that is not shared with other
actors. Communication and co-ordination between actors happens via message passing. 
The actor programming abstraction is a natural fit for distributed cloud
applications. Some of the popular instances of actor-based frameworks and
languages include Akka \cite{akka}, Erlang \cite{erlang}, and Orleans
\cite{bernstein2014orleans}.

For fault tolerance, each of these frameworks provide access to a
persistent store that is automatically restored to the last saved
state when a failed actor recovers. However, the responsibility of
committing the persistent state and ensuring that it is
consistent with the rest of the system still falls on the
programmer. Moreover, the communication
between actors is best-effort, and the programmer is again responsible
for managing retries and de-duplication of messages. More
specifically, unlike RSMs, these frameworks do not provide failure
transparency by-construction.

Orleans introduced the concept of ``virtual actors'': these actors need not be explicitly
created. They are instantiated on demand when they receive a message. Further, 
they are location independent, allowing the Orleans runtime
to dynamically load-balance the placement of actors across a
cluster, even putting frequently-communicating actors together \cite{actop}. 
RSM instances must be explicitly created, but they are location independent. 
Our implementation, however, currently does not attempt to move an RSM after it
has been created. The initial placement of a fresh RSM  can be controlled by the
programmer, after which the instance is permanently tied to that location.
Service Fabric Reliable Actors \cite{sf-actors} are also
an implementation of the virtual actors paradigm. We provide an empirical comparison of RSMs
with Reliable Actors in Sec.~\ref{Se:evalmicro}.

\noindent \textbf{Reactive programming.} Reactive frameworks \cite{DBLP:journals/csur/BainomugishaCCMM13} 
are used in the development of event-driven and interactive applications. These
frameworks provide a programmatic way of setting up a \textit{dataflow graph}
that marks functional dependencies between variables. As the value of certain
variables change over time, the rest of the dependent variables are updated
automatically. Recent work \cite{mogk2018fault} describes an extension to 
REScala \cite{DBLP:conf/aosd/SalvaneschiHM14} in order to provide
fault-tolerance support in distributed reactive programming. The
framework relies on taking snapshots of critical data and then uses replay to
construct the entire program state on failure. This requires deterministic
execution. Further, the input signals are not captured as part of the snapshots, 
causing them to differ on re-execution or even get duplicated. These issues require 
programmer support. On the other hand, RSMs can support non-deterministic handlers and guarantees
exact once processing because input (i.e. inbox) is part of the reliable state that RSMs
maintain. The REScala extension provides eventual consistency for updates to
shared data, making use of state-based conflict-free replicated data types
(CRDTs) \cite{shapiro2011comprehensive}. RSMs do not have shared state; maintaining
common state between two RSMs can be done by creating (and communicating with) another RSM that owns the
state. RSM messaging is reliable: this provides strong consistency between RSMs,
however, it is less resilient to network outages than CRDTs because the latter
allows for progress even in a disconnected state.

\noindent \textbf{Big-data analytics.} 
Big-data processing systems such as SPARK \cite{DBLP:conf/nsdi/ZahariaCDDMMFSS12} 
and SCOPE \cite{apolloOSDI2014} are popular frameworks for data analytics. They
provide a SQL-like programming interface that gets compiled to map-reduce
stages for distributed execution on a fault-tolerant runtime. These systems,
however, are meant for data-parallel batch processing. They execute on immutable input 
that is known ahead of time.   

\noindent \textbf{Other frameworks.} 
Ramalingam et al. \cite{DBLP:conf/popl/RamalingamV13} provide a monadic
framework that makes functional computation idempotent. Their transformation
records the sequence of steps that have already been executed. On
re-execution, such steps are skipped. Idempotent computation enables
fault-tolerance: simply keep re-executing until completion. Their work focuses
on state updates made by a single sequential agent. They assume determinism of
the computation and do not handle communication.
%%  \nitin{I'm not sure if this is 
%% accurate. Ramalingam's system does consist of several concurrent agents, like 
%% ours, which each run sequentially, like ours; but more importantly, though 
%% $\lambda_{FAIL}$ doesn't explicitly have non-determinism, they do briefly talk 
%% about how non-deterministic constructs can be treated as effectful steps 
%% and thus incorporated in their model. In fact, I think their system can do all 
%% that ours can, minus the P\# testing}
RSM programs, on
the other hand, support multiple concurrent agents with possible
non-deterministic execution. RSMs ensure idempotence by atomically committing
the effects of processing of a message along with the dequeue of the message
from the inbox.

Another class of languages for distributed systems, including Orca
\cite{DBLP:journals/tse/BalKT92} and X10
\cite{DBLP:conf/oopsla/CharlesGSDKEPS05}, rely on distributed shared memory. 
They enable applications that span multiple machines
while allowing the freedom to access memory across machine boundaries. They
mostly focus on in-memory computation, without  support for state
persistence or fault tolerance.
%============================================

\bibliographystyle{unsrt}  

\bibliography{theBibliography}

\end{document}